\documentclass{emulateapj}
\usepackage{amsmath}
\usepackage{amssymb}
\usepackage{epsfig}
\usepackage{subfigure}
\usepackage{natbib}
\usepackage{graphicx}
\usepackage{color}

\shorttitle{SN\,2010\MakeLowercase{ay}: a luminous broad-lined SN~Ic}
\shortauthors{Sanders et al.}

\newcommand{\GemEBmV}{0.2}
\newcommand{\GemAv}{0.6}
\newcommand{\GemPPohfourOthree}{8.19}
\newcommand{\GemPPohfoursol}{0.3}
\newcommand{\GemPPohfourNtwo}{8.26}
\newcommand{\GemKDohtwo}{8.50}
\newcommand{\GemZninefour}{8.49}
\newcommand{\GemRtwothree}{0.873\pm0.006}
\newcommand{\GemDirect}{8.24\pm0.08}
\newcommand{\GemTe}{1.09\pm0.06\times10^4}
\newcommand{\GemNe}{80\pm20}
\newcommand{\BalmerRatio}{3.45\pm0.02}
\newcommand{\Gemz}{0.06717}

\newcommand{\SDSSOHppohfour}{8.38}

\newcommand{\texp}{2010 February 21.3$\pm$1.3}

\newcommand{\texpthreepii}{4}

\newcommand{\Rpeak}{-20.2\pm0.2}

\newcommand{\Canosdev}{1.5}
\newcommand{\tpeak}{2010 March $18.5\pm 0.2$}

\newcommand{\tpeakDrakethree}{-1}
\newcommand{\UTDrakethree}{733848.4}
\newcommand{\MJDDrakethree}{55272.4}
\newcommand{\RmagDrakethree}{-20.4\pm0.2}
\newcommand{\tpeakDrakeone}{-29}
\newcommand{\UTDrakeone}{733820.4}
\newcommand{\MJDDrakeone}{55244.4}
\newcommand{\RmagDrakeone}{-19.2\pm0.2}
\newcommand{\tpeakDraketwo}{-13}
\newcommand{\UTDraketwo}{733836.4}
\newcommand{\MJDDraketwo}{55260.4}
\newcommand{\RmagDraketwo}{-19.3\pm0.2}
\newcommand{\tpeakPrieto}{4}
\newcommand{\UTPrieto}{733853.2}
\newcommand{\MJDPrieto}{55277.2}
\newcommand{\RmagPrieto}{-20.22\pm0.07}
\newcommand{\tpeakthreepii}{-21}
\newcommand{\UTthreepii}{733828.2}
\newcommand{\MJDthreepii}{55252.2}
\newcommand{\Rmagthreepii}{-16.8\pm0.3}
\newcommand{\tpeakthreepir}{-25}
\newcommand{\UTthreepir}{733824.6}
\newcommand{\MJDthreepir}{55248.6}
\newcommand{\Rmagthreepir}{-16.0\pm0.1}
\newcommand{\tpeakWHT}{14}
\newcommand{\UTWHT}{733863.0}
\newcommand{\MJDWHT}{55287.0}
\newcommand{\RmagWHT}{-19.2\pm0.2}
\newcommand{\tpeakGemini}{24}
\newcommand{\UTGemini}{733873.0}
\newcommand{\MJDGemini}{55297.0}
\newcommand{\RmagGemini}{-19.0\pm0.1}
\newcommand{\tpeakthreepirlate}{376}
\newcommand{\UTthreepirlate}{734225.0}
\newcommand{\MJDthreepirlate}{55649.0}
\newcommand{\Rmagthreepirlate}{-15.7\pm0.1}
\newcommand{\tpeakthreepiilate}{372}
\newcommand{\UTthreepiilate}{734221.0}
\newcommand{\MJDthreepiilate}{55645.0}
\newcommand{\Rmagthreepiilate}{-16.22\pm0.10}
\newcommand{\Gemfluxthreeseventwoseven}{52\pm1}
\newcommand{\Gemfluxfouronezerotwo}{4.5\pm0.2}
\newcommand{\Gemfluxfourthreefourzero}{9.9\pm0.1}
\newcommand{\Gemfluxfourthreesixthree}{0.7\pm0.1}

\newcommand{\Gemfluxfoureightsixone}{24.5\pm0.1}
\newcommand{\Gemfluxfourninefivenine}{30.4\pm0.1}
\newcommand{\Gemfluxfivezerozeroseven}{90.1\pm0.2}

\newcommand{\Gemfluxsixfivefoureight}{2.2\pm0.1}
\newcommand{\Gemfluxsixfivesixtwo}{84.5\pm0.1}
\newcommand{\Gemfluxsixfiveeightfour}{6.33\pm0.06}
\newcommand{\Gemfluxsixsevenoneseven}{7.46\pm0.06}
\newcommand{\Gemfluxsixseventhreeone}{5.59\pm0.06}
\newcommand{\VelGemTonrySiA}{18.3}

\newcommand{\VelGemTonryCa}{20.1}

\newcommand{\VelWhtTonrySiA}{19.2}

\newcommand{\VelWhtTonryCa}{21.7}

\newcommand{\Mni}{0.9^{+0.1}_{-0.1}}
\newcommand{\Mej}{4.7}
\newcommand{\Ek}{10.8}
\newcommand{\Mejbh}{6.1}
\newcommand{\Ekbh}{23.9}

\newcommand{\niejratio}{0.2}

\newcommand{\distance}{297.9}

\newcommand{\fmrOH}{8.20}

\newcommand{\SDSSOH}{$8.58^{+0.02}_{-0.03}$}
\newcommand{\SDSSMstar}{$3.6^{+2.9}_{-1.3} \times10^8~M_{\odot}$}
\newcommand{\SDSSSFR}{$1.0^{+0.3}_{-0.2}~M_{\odot}$~yr$^{-1}$}
\newcommand{\jhuAGNcutN}{167,837}
\newcommand{\jhuAGNcutMp}{4th}
\newcommand{\jhuAGNcutOHp}{11th}
\newcommand{\jhuAGNcutSFRp}{38th}

\newcommand{\jhuAGNMcutN}{6,978}
\newcommand{\jhuAGNMcutMup}{5.62}
\newcommand{\jhuAGNMcutMdown}{1.39}
\newcommand{\jhuAGNMcutOHp}{77th}

\newcommand{\jhuAGNMcutOHmed}{8.36}
\newcommand{\jhuAGNMcutSFRp}{96th}
\newcommand{\jhuAGNMcutSFRmed}{0.13}

\newcommand{\jhuAGNCcutN}{1,184}
\newcommand{\jhuAGNCcutURcol}{$0.47<u-r<0.67$}
\newcommand{\jhuAGNCcutOHp}{54th}
\newcommand{\jhuAGNCcutMp}{46th}
\newcommand{\jhuAGNCcutSFRp}{49th}
\newcommand{\HostAbsB}{$-18.35\pm0.05$}
\newcommand{\hostMBexpect}{$-15.8\pm1.3$}
\newcommand{\hostMBdiscrep}{2}

\newcommand{\jhuAGNMBcutSFRp}{26th}
\newcommand{\jhuAGNMBcutOHp}{3rd}
\newcommand{\jhuAGNMBcutOHmed}{$8.93\pm0.17$}

\newcommand{\KSEsfr}{1.1}

\newcommand{\vslopeef}{-0.8\pm0.4}
\newcommand{\vthirtyef}{6\pm2}
\newcommand{\vslopebw}{-0.86\pm0.08}
\newcommand{\vthirtybw}{10.3\pm0.7}
\newcommand{\vslopeap}{-1.5\pm0.7}
\newcommand{\vthirtyap}{5\pm2}
\newcommand{\vslopedh}{-0.9\pm0.1}
\newcommand{\vthirtydh}{12.0\pm0.9}
\newcommand{\vslopejd}{-0.5\pm0.2}
\newcommand{\vthirtyjd}{10\pm1}
\newcommand{\vslopelw}{-0.8\pm0.2}
\newcommand{\vthirtylw}{10\pm1}
\newcommand{\vslopeaj}{-0.3\pm0.2}
\newcommand{\vthirtyaj}{15\pm3}
\newcommand{\vslopebg}{-0.3\pm0.2}
\newcommand{\vthirtybg}{9\pm2}
\newcommand{\vslopebi}{-0.2\pm0.2}
\newcommand{\vthirtybi}{8\pm2}
\newcommand{\vsloperu}{-0.5\pm0.1}
\newcommand{\vthirtyru}{10.3\pm0.7}
\newcommand{\vslopebb}{-0.84\pm0.08}
\newcommand{\vthirtybb}{10.7\pm0.4}
\newcommand{\vslopeay}{-0.4\pm0.3}
\newcommand{\vthirtyay}{21\pm2}
\newcommand{\vslopebh}{-0.22\pm0.07}
\newcommand{\vthirtybh}{24\pm3}

\newcommand{\lcchi}{0.3}

\newcommand{\velfitslopegrbmean}{-0.7}
\newcommand{\velfitslopegrbstd}{0.2}
\newcommand{\velfitslopeicblmean}{-0.8}
\newcommand{\velfitslopeicblstd}{0.4}

\newcommand{\gps}{\ensuremath{g_{\rm P1}}}
\newcommand{\rps}{\ensuremath{r_{\rm P1}}}
\newcommand{\ips}{\ensuremath{i_{\rm P1}}}
\newcommand{\zps}{\ensuremath{z_{\rm P1}}}
\newcommand{\yps}{\ensuremath{y_{\rm P1}}}

\newcommand{\PS}{\protect \hbox {Pan-STARRS1}}
\def\amin{\char'023 }
\def\asec{\char'175 }

\def\na{\rm{New A}} 
\def\actaa{\rm{Acta Astron.}}   
\def\vSi{v_{\rm Si}}

\begin{document}

\title{SN\,2010ay is a Luminous and Broad-lined Type Ic Supernova within a Low-metallicity Host Galaxy}
\author{N. E. Sanders\altaffilmark{1}, 
A. M. Soderberg\altaffilmark{1}, 
S. Valenti\altaffilmark{2}, 
R. J. Foley\altaffilmark{1}, 
R. Chornock\altaffilmark{1}, 
L. Chomiuk\altaffilmark{1,3}, 
E. Berger\altaffilmark{1}, 
S. Smartt\altaffilmark{2}, 
K. Hurley\altaffilmark{4}, 
S. D. Barthelmy\altaffilmark{5}, 
E. M. Levesque\altaffilmark{6}, 
G. Narayan\altaffilmark{7}, 
M.T. Botticella\altaffilmark{2}, 
M. S. Briggs\altaffilmark{8}, 
V. Connaughton\altaffilmark{8}, 
Y. Terada\altaffilmark{9}, 
N. Gehrels\altaffilmark{5}, 
S. Golenetskii\altaffilmark{10}, 
E. Mazets\altaffilmark{10}, 
T. Cline\altaffilmark{11}, 
A. von Kienlin\altaffilmark{12}, 
W. Boynton\altaffilmark{13}, 
K. C. Chambers\altaffilmark{14}, 
T. Grav\altaffilmark{15}, 
J. N. Heasley\altaffilmark{14}, 
K. W. Hodapp\altaffilmark{14}, 
R. Jedicke\altaffilmark{14}, 
N. Kaiser\altaffilmark{14}, 
R. P. Kirshner\altaffilmark{1}, 
R.-P. Kudritzki\altaffilmark{14}, 
G. A. Luppino\altaffilmark{14}, 
R. H. Lupton\altaffilmark{16}, 
E. A. Magnier\altaffilmark{14}, 
D. G. Monet\altaffilmark{17}, 
J. S. Morgan\altaffilmark{14}, 
P. M. Onaka\altaffilmark{14}, 
P. A. Price\altaffilmark{16}, 
C. W. Stubbs\altaffilmark{7}, 
J. L. Tonry\altaffilmark{14}, 
R. J. Wainscoat\altaffilmark{14}, 
M. F. Waterson\altaffilmark{18}. }

\altaffiltext{1}{Harvard-Smithsonian Center for Astrophysics, 60 Garden Street, Cambridge, MA 02138 USA}
\altaffiltext{2}{Astrophysics Research Centre, School of Maths and Physics,Queen’s University, BT7 1NN, Belfast, UK}
\altaffiltext{3}{National Radio Astronomy Observatory, Socorro, NM 87801, USA}
\altaffiltext{4}{Space Sciences Laboratory, University of California Berkeley, 7 Gauss Way, Berkeley, CA 94720, USA}
\altaffiltext{5}{NASA Goddard Space Flight Center, Greenbelt, MD 20771, USA}
\altaffiltext{6}{CASA, Department of Astrophysical and Planetary Sciences, University of Colorado, 389-UCB, Boulder, CO 80309, USA}
\altaffiltext{7}{Department of Physics, Harvard University, Cambridge, MA 02138, USA} 
\altaffiltext{8}{CSPAR, University of Alabama in Huntsville, Huntsville, Alabama, USA.} 
\altaffiltext{9}{Department of Physics, Saitama University, Shimo-Okubo, Sakura-ku, Saitama-shi, Saitama 338-8570} 
\altaffiltext{10}{Ioffe Physico-Technical Institute, Laboratory for Experimental Astrophysics, 26 Polytekhnicheskaya, St. Petersburg 194021, Russia} 
\altaffiltext{11}{Emeritus, NASA Goddard Space Flight Center, Code 661, Greenbelt, MD 20771, USA} 
\altaffiltext{12}{Max-Planck Institut f\"{u}r extraterrestrische Physik, 85748 Garching, Germany}
\altaffiltext{13}{Lunar and Planetary Laboratory, University of Arizona, Tucson, AZ 85721, USA}
\altaffiltext{14}{Institute for Astronomy, University of Hawaii at Manoa, Honolulu, HI 96822, USA} 
\altaffiltext{15}{Department of Physics and Astronomy, Johns Hopkins University, 3400 North Charles Street, Baltimore, MD 21218, USA} 
\altaffiltext{16}{Department of Astrophysical Sciences, Princeton University, Princeton, NJ 08544, USA} 
\altaffiltext{17}{US Naval Observatory, Flagstaff Station, Flagstaff, AZ 86001, USA} 
\altaffiltext{18}{International Center for Radio Astronomy Research, The University of Western Australia, Crawley, Perth, Australia }

\email{nsanders@cfa.harvard.edu}

\begin{abstract}

We report on our serendipitous pre-discovery detection and detailed follow-up of the broad-lined Type Ic supernova (SN) 2010ay at $z=0.067$ imaged by the \PS\ $3\pi$ survey just $\sim\texpthreepii{}$ days after explosion. The SN had a peak luminosity, $M_R\approx -20.2$~mag, significantly more luminous than known GRB-SNe and one of the most luminous SNe Ib/c ever discovered. The absorption velocity of SN\,2010ay is $\vSi\approx19\times10^3~$~km~s$^{-1}$ at $\sim40$~days after explosion, $2-5$ times higher than other broad-lined SNe and similar to the GRB-SN\,2010bh at comparable epochs. Moreover, the velocity declines $\sim2$ times slower than other SNe~Ic-BL and GRB-SNe. Assuming that the optical emission is powered by radioactive decay, the peak magnitude implies the synthesis of an unusually large mass of $^{56}$Ni, $M_{\rm Ni}=0.9~M_\odot$. Modeling of the light-curve points to a total ejecta mass, $M_{\rm ej}\approx\Mej{} M_{\odot}$, and total kinetic energy, $E_{K}\approx 11\times10^{51}$~ergs. The ratio of $M_{\rm Ni}$ to $M_{\rm ej}$ is $\sim2$ times as large for SN\,2010ay as typical GRB-SNe and may suggest an additional energy reservoir. The metallicity ($\log({\rm O/H})_{\rm PP04}+12=\GemPPohfourOthree{}$) of the explosion site within the host galaxy places SN\,2010ay in the low-metallicity regime populated by GRB-SNe, and $\sim0.5(0.2)$ dex lower than that typically measured for the host environments of normal (broad-lined) Ic supernovae. We constrain any gamma-ray emission with $E_\gamma\lesssim6\times10^{48}$~erg (25-150 keV) and our deep radio follow-up observations with the Expanded Very Large Array rule out relativistic ejecta with energy, $E\gtrsim 10^{48}$~erg. We therefore rule out the association of a relativistic outflow like those which accompanied SN\,1998bw and traditional long-duration GRBs, but place less-stringent constraints on a weak afterglow like that seen from XRF\,060218. These observations challenge the importance of progenitor metallicity for the production of a GRB, and suggest that other parameters also play a key role.

\smallskip
\end{abstract}


\section{INTRODUCTION}
\label{sec:intro} 

Recent observations have shown that long-duration gamma-ray bursts are
accompanied by Type Ic supernovae (SNe) with broad absorption features
(hereafter, ``broad-lined,'' BL), indicative of high photospheric
expansion velocities \citep[see][for a review]{woosley06}.  This
GRB-SN connection is popularly explained by the favored ``collapsar
model'' \citep{MWH01} in which the gravitational collapse of a massive
($M\gtrsim 20~M_{\odot}$) progenitor star gives birth to a central
engine -- a rapidly rotating and accreting compact object -- that
powers a relativistic outflow.  At the same time, not all SNe~Ic-BL
show evidence for a central engine. Radio observations constrain
the fraction of SNe~Ic-BL harboring relativistic outflows to be less
than a third
\citep{skn+06,scp+10}.

The physical parameter(s) that distinguish the progenitors of
GRB-associated SNe from other SNe~Ic-BL remains debated, while
theoretical considerations indicate that progenitor metallicity may
play a key role \citep{WoosleyHeger}. In the collapsar scenario,
massive progenitor stars with metallicity above a threshold,
$Z\gtrsim0.3~Z_\odot$, lose angular momentum to metal line-driven
winds, preventing the formation of a rapidly rotating compact remnant,
and in turn, a relativistic outflow.  At the same time, the
hydrogen-free spectra of SNe~Ic-BL indicate that their stellar
envelopes have been stripped prior to explosion, requiring higher
metallicities (e.g.,~$Z\approx Z_{\odot}$) if due to radiation
driven-winds \citep{Woosley95}.  Alternatively, short-period
($\sim0.1$~days) binary interaction may be invoked to spin up stars
via tidal forces as well as cause mass loss via Roche lobe overflow
\citep{Podsiadlowski04,Fryer05}. However, even in the binary scenario,
GRB formation is predicted to occur at higher rates in
lower-metallicity environments, where the radius and mass loss rates
of stars should be smaller \citep{Izzard04}.

Observationally, {\it most} GRB-SNe are discovered within dwarf
star-forming galaxies \citep{Fruchter06} characterized by sub-solar
metallicities, $Z\lesssim0.5~Z_\odot$ \citep{Levesque10}.  This has
been interpreted as observational support for the
metallicity-dependent collapsar model \citep{Stanek06}. Meanwhile, SNe~Ic-BL without
associated GRBs have historically been found in more enriched
environments \citep{Modjaz08}, motivating the suggestion of an
observationally determined ``cut-off metallicity'' above which GRB-SNe
do not form \citep{Kocevski09}.  However, this difference may be
partly attributable to the different survey strategies: SNe have been
found in large numbers by galaxy-targeted surveys biased towards more
luminous (and therefore higher metallicity) environments, while GRB
host galaxies are found in an untargeted manner through their
gamma-ray emission.

Against this backdrop of progress, recent observations have begun to
call into question some aspects of this scenario. First, several
long-duration GRBs have now been identified in solar or super-solar
metallicity environments (e.g., GRB\,020819; \citealt{lkg+10}).
Similarly, the luminous radio emission seen from
SN~Ic-BL 2009bb pointed unequivocally to the production of copious
relativistic ejecta resembling a GRB afterglow
\citep{scp+10} while the explosion environment 
was characterized by a super-solar metallicity, $Z\sim 1-2~Z_{\odot}$
\citep{lsf+10}. Together with the growing lack of evidence for massive 
progenitor stars for SNe~Ic in pre-explosion {\it Hubble Space
Telescope} images \citep{smartt09}, a lower mass ($M\sim
10-20~M_{\odot}$) binary progenitor system model (with a gentler
metallicity dependence) is gaining increasing popularity
\citep{ywl10}.  Multi-wavelength studies of SNe~Ic-BL in metal-poor 
environments may shed further light on the role of metallicity in the
nature of the progenitor and the explosion properties, including the
production of relativistic ejecta.

Fortunately, with the recent advent of wide-field optical transient
surveys (e.g., Catalina Real-time Transient Survey (CRTS);
\citealt{CSS}, Panchromatic Survey Telescope and Rapid Response System; (\PS, abbreviated PS1) \citealt{PS1}, Palomar Transient Factory;
\citealt{PTF})  SNe~Ic-BL are now being discovered in metal-poor
environments, $Z\sim 0.5~Z_{\odot}$
\citep{Arcavi10,Modjaz10} thanks to an unbiased 
search technique.  In this paper, we present pre-discovery \PS\
imaging and multi-wavelength follow-up observations for the broad-lined Type~Ic SN\,2010ay discovered by CRTS (\citealt{DrakeCBET}). In \S\ref{sec:obs}, we report our optical (\PS, Gemini, William Herschel Telescope) and radio (Expanded Very Large Array; EVLA)
observations. In \S\ref{res:snprop}, we model the optical light-curve
and analyze the spectra to derive the explosion properties of
SN\,2010ay. In \S\ref{res:evla}, we use our observations of
SN\,2010ay with the EVLA to place strict limits on the presence of
relativistic outflow.  In \S\ref{res:noGRB}, we draw from gamma-ray
satellite coverage to rule out a {\em detected} gamma-ray burst in
association with SN\,2010ay.  In \S\ref{res:metallicity}, we derive
the explosion site metallicity and find it to be significantly
sub-solar and typical of most GRB-SNe host environments. In
\S\ref{res:phys}, we discuss the implications of these findings in the
context of the explosion and progenitor properties and conclude in
\S\ref{sec:conc}.

\section{OBSERVATIONS}
\label{sec:obs}

\subsection{Discovery by CRTS} \label{obs:phot}

SN\,2010ay was discovered by the Catalina Real-time Transient Survey \citep{CSS} on 2010
March 17.38 UT \citep{DrakeCBET} and designated
CSS100317:123527+270403, with an unfiltered magnitude of
$m\approx17.5$~mag and located $\lesssim1$'' of the center of a
compact galaxy, \object{SDSS J123527.19+270402.7} at $z=0.067$ (Table
\ref{tab:SDSS}). We adopt a distance, $D_L=\distance{}$ Mpc, to the host
galaxy\footnote{We assume $H_0=71$~km~s$^{-1}$ Mpc$^{-1}$,
$\Omega_\Lambda = 0.73, \Omega_M=0.27$}, and note that the Galactic
extinction toward this galaxy is $E(B-V)=0.017$ \citep{SFD}. Pre-discovery
unfiltered images from CRTS revealed an earlier detection of
the SN on Mar 5.45 UT at $m\approx 18.2$~mag and a
non-detection from Feb 17.45 UT at $m
\gtrsim 18.3$~mag
\citep{DrakeCBET}. 

A spectrum obtained on Mar 22 UT revealed the SN to be of Type Ic with
broad features, similar to the GRB-associated SN 1998bw spectrum
obtained near maximum light \citep{FilippenkoCBET}. This
classification was confirmed by \cite{PrietoCBET} who additionally
reported photometry for the SN (see Table~\ref{tab:phot}). After
numerically subtracting the host galaxy emission, they estimate an
unusually luminius absolute magnitude of $V\approx-19.4$~mag.

\begin{deluxetable}{lc}
\tablecaption{SN\,2010ay host galaxy SDSS J123527.19+270402.7\label{tab:SDSS}}
\tablewidth{0pt}
\tabletypesize{\scriptsize}
\tablehead{ \colhead{Parameter} & \colhead{Value}}
\startdata
RA & 12$^{\rm h}$35$^{\rm m}$27$^{\rm s}$.19 (J2000) \\
DEC & +27$^{\circ}$04$\amin$02.7\asec\ (J2000) \\
Redshift (z) & $0.0671$ \\
Petrosian radius & $1.355$\asec\ \\
\cutinhead{Photometry\tablenotemark{a}}
$u'$ & $19.56 \pm 0.03$~mag\\
$g'$ & $19.02 \pm 0.01$~mag\\
$r'$ & $19.02 \pm 0.01$~mag\\
$i'$ & $18.69 \pm 0.01$~mag\\
$z'$ & $18.87 \pm 0.04$~mag\\
$U$ & $19.50\pm0.06$~mag\\
$B$ & $19.02\pm0.05$~mag\\
$V$ & $19.13\pm0.05$~mag\\
$R$ & $18.94\pm0.05$~mag\\
$I$ & $18.90\pm0.06$~mag\\
\cutinhead{Extinction}
E(B-V)$_{\rm MW}$\tablenotemark{b} & 0.017~mag\\  
E(B-V)$_{\rm host}$\tablenotemark{c} & 0.2~mag\\
\enddata
\tablecomments{SDSS host galaxy properties and $ugriz$ photometry.}
\tablenotetext{a}{Model magnitudes from SDSS DR8 \citep{sdss8}. Host galaxy photometry has not been corrected for extinction. $UBVRI$ photometry has been converted from the SDSS $ugriz$ measurements using the transformation of \cite{Blanton07}.}
\tablenotetext{b}{The Milky Way extinction as determined by \cite{SFD}, assuming $R_V=3.1$.}
\tablenotetext{c}{The host galaxy extinction determined from the SDSS spectrum centered on the galaxy nucleus, via the Balmer decrement as described in \S\ref{obs:host}.}
\end{deluxetable}

\subsection{Pre-Discovery Detection With \PS\ $3\pi$} \label{obs:photps1}

The field of SN\,2010ay was serendipitously observed with the
PS1 $3\pi$ survey in the weeks preceding its discovery.
\PS\ is a wide-field imaging system at Haleakala, 
Hawaii dedicated to survey observations \citep{PS1}. The PS1 optical
design \citep{PS1opt} uses a 1.8~meter diameter $f$/4.4 primary
mirror, and a 0.9~m secondary. The telescope illuminates a diameter of
3.3 degrees. The \PS\ imager \citep{PS1cam} comprises a total
of 60 $4800\times4800$ pixel detectors, with 10~$\mu$m pixels that
subtend 0.258~arcsec. The PS1 observations are obtained through a set
of five broadband filters designated as \gps, \rps,
\ips, \zps, and \yps. These filters are similar to those used in
previous surveys, such as SDSS \citep{SDSSfilter}. However, The \gps\
filter extends 20~nm redward of $g'$, the \zps\ filter is cut off at
930~nm, and SDSS has no corresponding \yps\ filter \citep{PS1cal}.

\begin{figure*}
\includegraphics[angle=0,scale=.8]{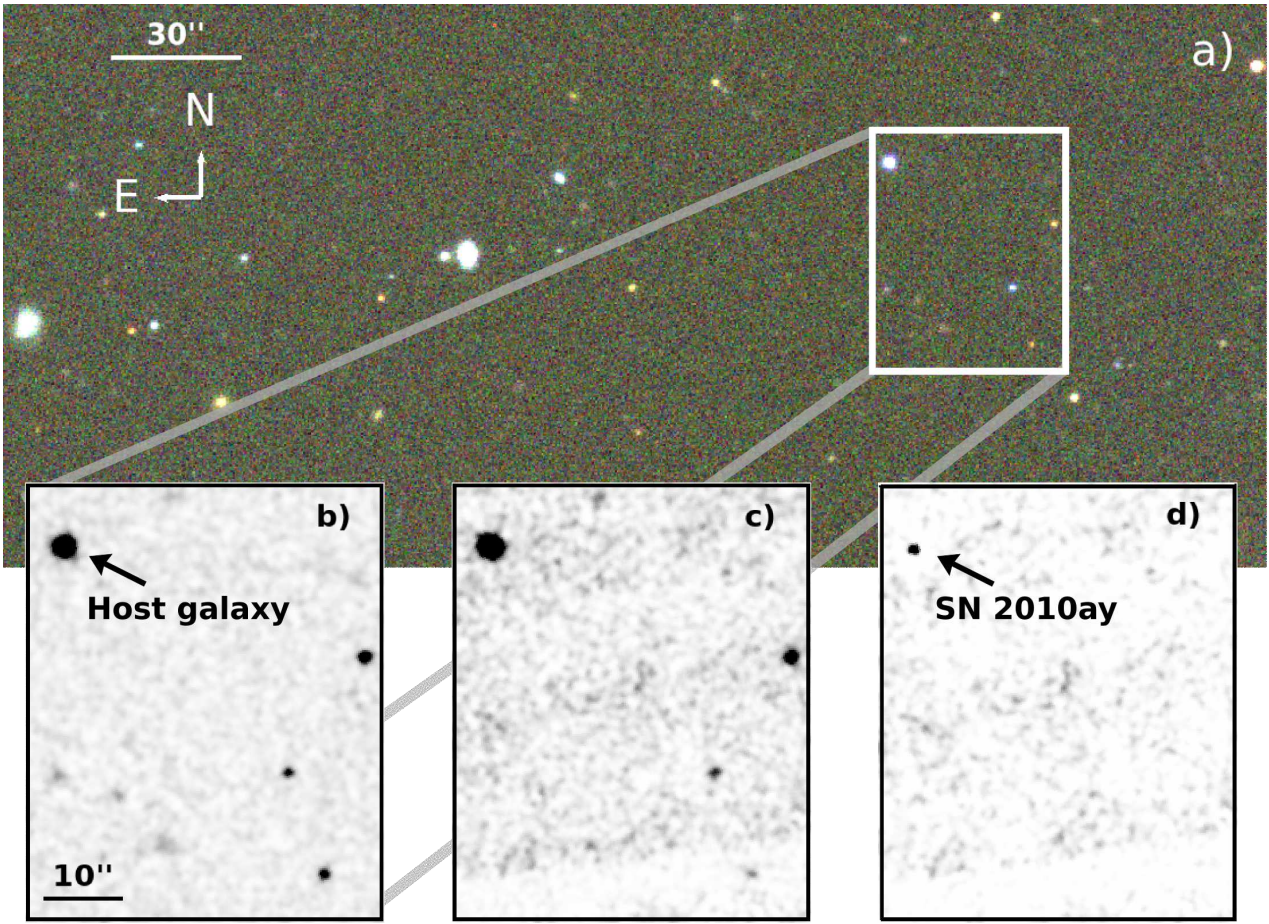}
\caption{Images illustrating the \PS\ pre-discovery detection of SN\,2010ay and the surrounding field. a) Pre-explosion $gri$-composite image from the SDSS DR7 \citep{sdss7}, observed 2004 December 21, b) SDSS $i$-band image geometrically registered to the PS1 image frames and zoomed in to the host galaxy of SN 2010ay, c) \ips-band image from the $3\pi$ survey of PS1, observed 2010 February 25, d) the difference of the SDSS $i'$ and PS1 \ips{} images. The transient emission can be seen in the NE corner of the last panel.  Nearby stars are included in the zoomed-in frames to illustrate the efficiency of subtraction.\label{fig:ps1}}
\end{figure*}

The field of SN\,2010ay was observed on 2010 February 21st (\rps-band)
and February 25th (\ips-band, Figure \ref{fig:ps1}). On each night, four exposures were
collected following the strategy of the PS1-3pi survey
(Chambers et al., in preparation). Following the CRTS discovery and announcement of SN\,2010ay, 
we geometrically registered SDSS pre-explosion images to the PS1
images and performed digital image subtraction using the {\it ISIS}
package \citep{Alard00}. No residual flux was found in the difference
\rps-band image from 2010 February 21 with an upper limit of $r > 
22.0$ mag. However, we detect residual flux at the position of
SN\,2010ay in the \ips-band residual image from 2010 February 25 with
a magnitude of $i′ = 21.1 \pm 0.3$ mag.

The field was again observed in the \ips\-band on 2011 March 25 and
the \rps-band on 2011 March 29, but no residual flux was detected in
the subtractions at the SN position to limits of $i'\gtrsim 22.2$ and
$r'\gtrsim 21.9$.

\begin{figure}
\plotone{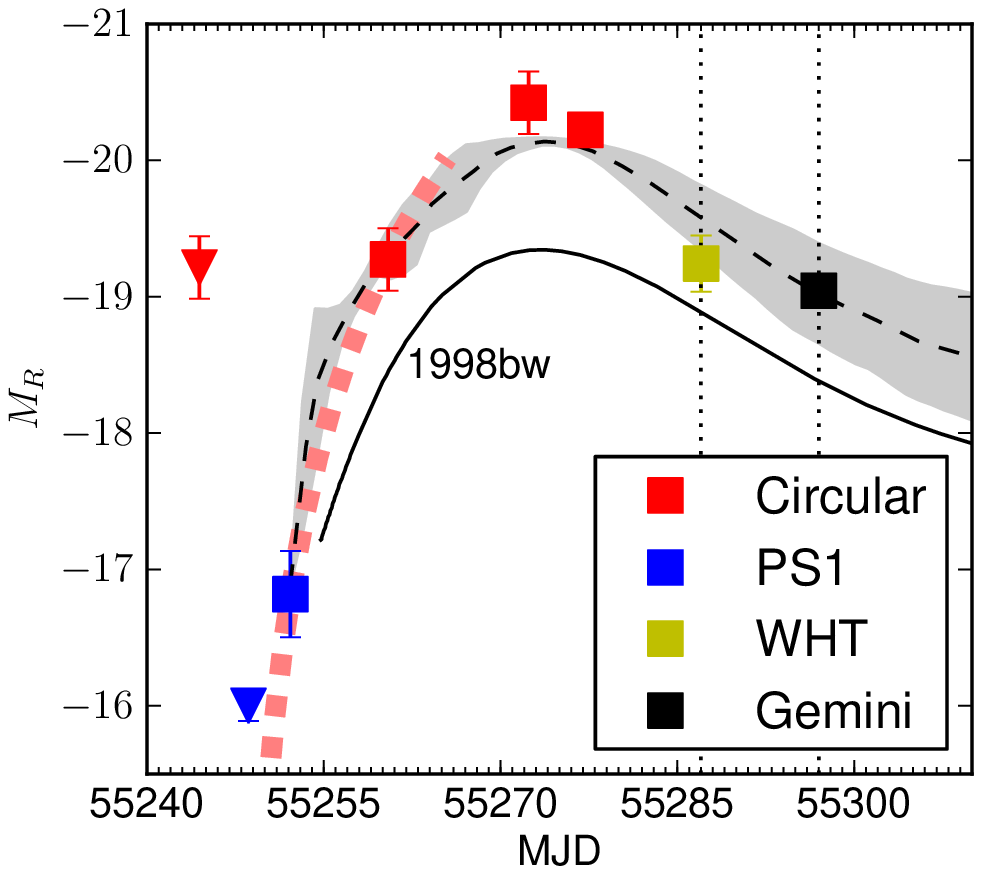}
\caption{The optical R-band light curve of SN\,2010ay, as compiled in Table \ref{tab:phot} from CBET 2224 (red squares), the PS1 $3\pi$ survey (blue), our Gemini images (black), and synthetic photometry (\S\ref{obs:opt}) based on our WHT spectrum (gold). Triangles denote upper limits. The thick dashed line represents the luminosity of an expanding fireball fit to our early-time photometry (\S\ref{exp:lcmodel}). The thin dashed line is the SN Ib/c light-curve template of \cite{Drout10} and the gray field represents the standard deviation among its constituent photometry. The template is stretched by $(1+z)=1.067$ with the best fit parameters $t_{{\rm R peak}}={\rm March} 18\pm 2$ (MJD $55273\pm 2$), $M_{{\rm R peak}}=\Rpeak{}$~mag. The solid line is the light-curve of the GRB-SN 1998bw \citep{Galama98}, redshifted to match SN\,2010ay. The vertical dotted lines mark the epochs of our Gemini and WHT spectroscopy.\label{fig:lc}}
\end{figure}

In Table \ref{tab:phot} and Figure \ref{fig:lc}, we compile photometry
from the PS1 detections, our optical observations, and the circulars
to construct a light-curve for SN\,2010ay.

\begin{deluxetable*}{lp{35pt}lllrr}
\tablecaption{SN\,2010ay light-curve photometry\label{tab:phot}}
\tablewidth{0pt}
\tabletypesize{\scriptsize}
\tablehead{\colhead{UT date} & \colhead{MJD} & \colhead{$t_{peak}$\tablenotemark{a}} & \colhead{Filter}& \colhead{m\tablenotemark{b}}& \colhead{$M_R$\tablenotemark{c}} & \colhead{Source}}
\startdata
\UTDrakeone{}          & \MJDDrakeone{} & \tpeakDrakeone{} & \nodata & $<18.3$& $<\RmagDrakeone{}$ & \tablenotemark{d} \\ 
\UTthreepir{}          & \MJDthreepir{} & \tpeakthreepir{} & $\rps$     & $<22.0\pm0.1$ & $<\Rmagthreepir{}$ & \tablenotemark{e} \\
\UTthreepii{}          & \MJDthreepii{} & \tpeakthreepii{} & $\ips$     & $21.1\pm0.3$ & $\Rmagthreepii{}$& \tablenotemark{e} \\
\UTDraketwo{}          & \MJDDraketwo{}& \tpeakDraketwo{} & \nodata & $18.2$& $\RmagDraketwo{}$ & \tablenotemark{d}\\
\UTDrakethree{}        & \MJDDrakethree{} & \tpeakDrakethree{} & \nodata & $17.5$& $\RmagDrakethree{}$ & \tablenotemark{d} \\
\UTPrieto{}            & \MJDPrieto{} & \tpeakPrieto{} & B     & $18.39 \pm 0.05$ & \nodata& \tablenotemark{f} \\ 
\UTPrieto{}            & \MJDPrieto{} & \tpeakPrieto{} & V     & $17.61 \pm 0.05$ & \nodata& \tablenotemark{f} \\ 
\UTPrieto{}            & \MJDPrieto{} & \tpeakPrieto{} & R     & $17.44 \pm 0.05$ & $\RmagPrieto$& \tablenotemark{f} \\ 
\UTWHT{}               & \MJDWHT{} & \tpeakWHT{} & R     & $18.2\pm0.2$& $\RmagWHT{}$ & \tablenotemark{g} \\
\UTGemini{}            & \MJDGemini{} & \tpeakGemini{} & g     & $18.9\pm0.1$ & \nodata & \tablenotemark{h} \\
\UTGemini{}            & \MJDGemini{} & \tpeakGemini{} & r     & $18.3\pm0.1$& $\RmagGemini{}$ & \tablenotemark{h}\\
\UTGemini{}            & \MJDGemini{} & \tpeakGemini{} & i     & $18.0\pm0.1$ & \nodata & \tablenotemark{h} \\ 
\UTthreepiilate{}      & \MJDthreepiilate{} & \tpeakthreepiilate{} & $\ips$ & $<22.2\pm0.2$& $<\Rmagthreepiilate$ & \tablenotemark{e} \\
\UTthreepirlate{}      & \MJDthreepirlate{} & \tpeakthreepirlate{} & $\rps$ & $<21.9\pm0.1$& $<\Rmagthreepirlate$ & \tablenotemark{e} 
\enddata
\tablenotetext{a}{Time since peak in days, relative to the fitted value: \tpeak{}.}
\tablenotetext{b}{The measured apparent magnitude of the source, in the filter noted and without extinction correction. For the \PS\ photometry, a template image was subtracted; for the other points, the host galaxy flux has not been subtracted.}
\tablenotetext{c}{The absolute R magnitude of the SN. Filter conversion, host flux subtraction, and extinction correction have been performed as described in \S\ref{exp:lcmodel}.}
\tablenotetext{d}{\cite{DrakeCBET}, unfiltered (synthetic V-band).}
\tablenotetext{e}{Photometry from the \PS\ $3\pi$ survey.}
\tablenotetext{f}{\cite{PrietoCBET}.} 
\tablenotetext{g}{Synthetic photometry obtained from our WHT spectrum as described in \S\ref{obs:opt}.}
\tablenotetext{h}{Photometry from our Gemini/GMOS observations described in \S\ref{obs:photps1}.}
\end{deluxetable*}

\subsection{Optical observations} \label{obs:opt}

We obtained an optical spectrum ($\sim 3000-11000~\rm \AA$) of SN\,2010ay
on April 1 UT, from the ISIS blue arm instrument of the 4.2 m William
Herschel Telescope (WHT) at the Roque de Los Muchachos Observatory.
The spectrum was taken at the parallactic angle and the exposure time
was 1800~sec. We obtained a second 1800~sec optical spectrum ($\sim
3600-9600~\rm \AA$) using the Gemini Multi-Object Spectrograph (GMOS) on
the 8.1 m Gemini North telescope on 2010 April 11.4 UT. We employed
standard two-dimensional long-slit image reduction and spectral
extraction routines in IRAF\footnote{IRAF is distributed by the
National Optical Astronomy Observatory, which is operated by the
Association of Universities for Research in Astronomy (AURA) under
cooperative agreement with the National Science Foundation.}. We do
not apply a correction for atmospheric differential refraction,
because the displacement should be $\lesssim0.5$\asec\ at the airmass
of the observations, $\approx 1.0$.

\begin{figure*}
\plotone{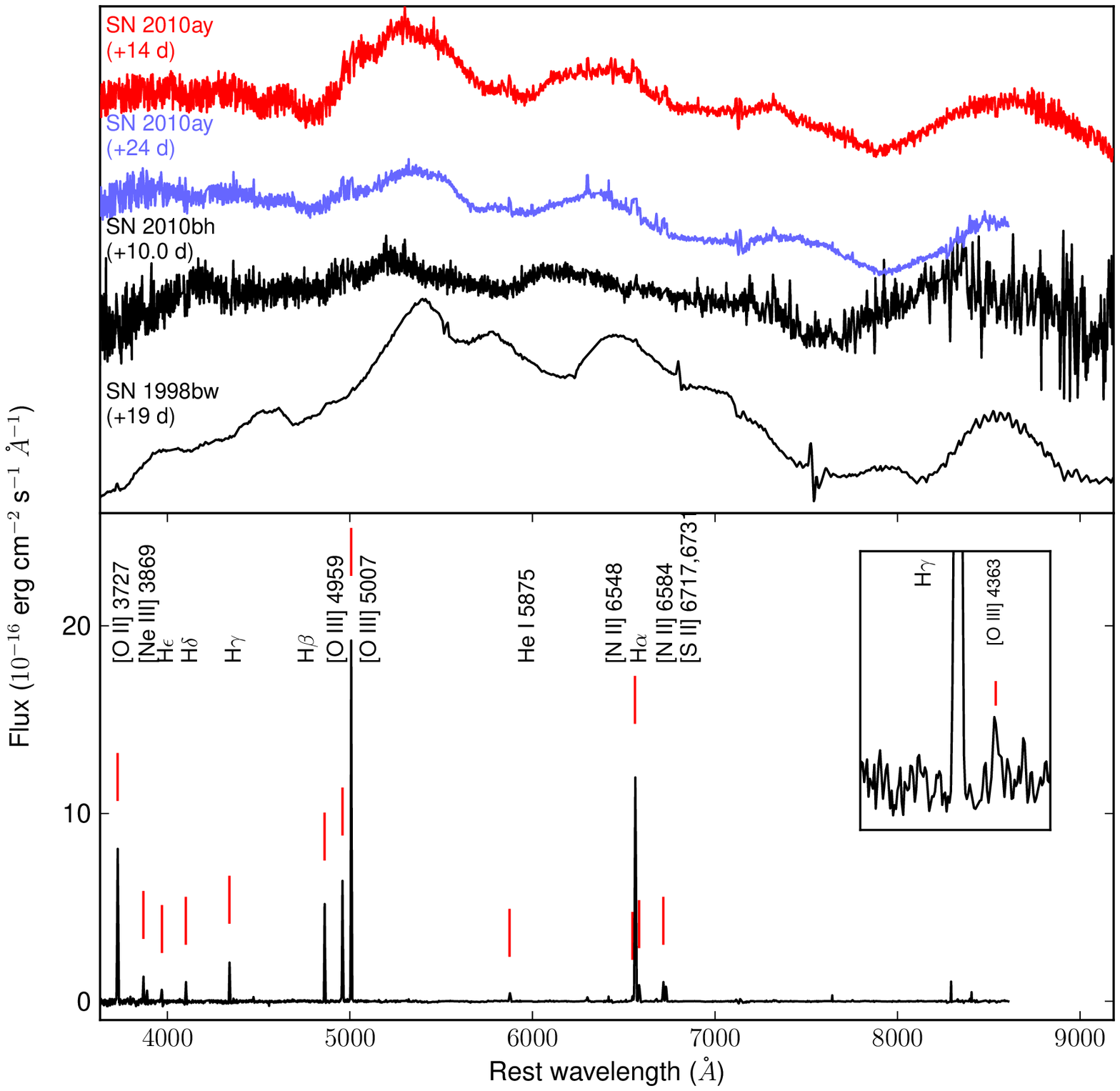}
\caption{Optical spectra of SN\,2010ay from Gemini (\tpeakGemini{} days after R-band peak) and the WHT (\tpeakWHT{} days). The spectrum is plotted decomposed into SN (above, with narrow lines clipped) and host galaxy (below, from Gemini, with spline-fit subtracted) components for clarity. The spectrum of SN\,2010bh from \cite{Chornock2010bh} is given in black for comparison, at 21.2 days after the GRB 100316D trigger \citep[$\sim10.0$ days after R-band peak,][]{Cano11}. The spectrum of SN 1998bw at +19 days from \cite{Patat01} is also plotted. Both are transformed to the redshift of SN\,2010ay. The SNe are shifted in flux for clarity. In the lower plot, relevant host galaxy emission lines are marked with a red line and labeled.\label{fig:spectra}}
\end{figure*}

In both our Gemini and WHT spectra, broad absorption features
associated with the SN are clearly detected in addition to narrow
emission lines typical of star forming galaxies. We distinguish the
host galaxy emission from the continuum dominated by the highly broadened SN emission by
subtracting a high-order spline fit to the continuum. Both SN and host galaxy spectral
components are shown in Figure~\ref{fig:spectra}. As illustrated in
the Figure, the broad, highly-blended spectral features of SN\,2010ay
resemble those of the Type Ic-BL SN\,2010bh (associated with GRB100316D) at a similar similar epoch
\citep{Chornock2010bh}. In particular, the broadening and blueshift
of the feature near 6355~$\rm \rm \AA$ are similar for SN\,2010ay and SN\,2010bh, and are broader and more blueshifted
than in SN\,1998bw at a comparable epoch.  We discuss the comparison between these two SN further in \S\ref{obs:vel} and \S\ref{host:compare}.

Additionally, we obtained 60 s $gri$-band images of SN\,2010ay on 2010 April 11.4 UT using GMOS. The data were reduced using the {\tt gemini} package in IRAF, and photometry was performed using the standard GMOS zero-points\footnote{http://www.gemini.edu/?q=node/10445}. We measure that $[g,r,i]=[18.90,18.32,18.04]\pm0.1$~mag.

Imaging photometry is not available at the epoch of our WHT spectrum.  However, the spectrum was flux-calibrated against observations of the standard star Feige34, which was observed the same night and at approximately the same low airmass as the supernova.  For the observations of both the standard star and supernova, the slit was placed at the parallactic angle.
We perform synthetic photometry on the spectrum to extract the flux at the central frequency of the R-band ($6527~\rm \AA$) and
find $R=18.2\pm0.2$ mag.  We then subtract the host galaxy flux numerically.

\subsubsection{Host galaxy features} \label{obs:host}

We measure fluxes of the narrow emission lines from the host galaxy in
our Gemini spectrum, as reported in Table \ref{tab:lines}. We fit a
Gaussian profile to each narrow emission line; for nearby lines such
as [\ion{N}{2}] and H$\alpha$, we fit multiple Gaussians
simultaneously. We model the local continuum with a linear fit to
$20~\rm \AA$ regions on either side of each line.  We estimate uncertainties in quantities derived from the line fluxes by Monte Carlo propagation of the uncertainties in the flux measurement.

The host galaxy is significantly reddened as evidenced by the flux ratio of
H$_\alpha$ to H$_\beta$, $\approx \BalmerRatio{}$.  We infer
${\rm E(B-V)}=\GemEBmV{}$~mag ($A_V=\GemAv{}$~mag), as measured from the
Balmer decrement in our Gemini spectrum, assuming $R_V=3.1$, Case B
recombination \citep{oandf}, and the reddening law of
\cite{cardelli89}. This is similar to the value derived
from the SDSS DR8 nuclear fiber spectroscopy line fluxes for the host
galaxy (${\rm E(B-V)}=0.2$~mag). The value reported in \cite{ModjazATel} was
also similar: E(B-V)$=0.3$~mag. Furthermore, we note that the 
color ($B-V=0.78\pm0.07$~mag) as reported by \cite{PrietoCBET} at \tpeakPrieto{}~d after
$R-$band peak is significantly redder than SN Ib/c color curve
templates \cite{Drout10}, further supporting a non-negligible host
galaxy extinction.

\begin{deluxetable}{lr}
\tablecaption{Emission line fluxes measured for the host galaxy of SN\,2010ay\label{tab:lines}}
\tablewidth{0pt}
\tabletypesize{\scriptsize}
\tablehead{ \colhead{Emission Line} & \colhead{Flux} \\ & ($10^{-16}$erg ~s$^{-1}$cm$^{-2}$)}
\startdata
[\ion{O}{2}]$\lambda3726,3729$   & $\Gemfluxthreeseventwoseven{}$ \\
H$\delta$          & $\Gemfluxfouronezerotwo{}$   \\
H$\gamma$          & $\Gemfluxfourthreefourzero{}$ \\
{[\ion{O}{3}]}$\lambda4363$     & $\Gemfluxfourthreesixthree{}$ \\
H$\beta$           & $\Gemfluxfoureightsixone{}$  \\
{[\ion{O}{3}]}$\lambda4959$     & $\Gemfluxfourninefivenine{}$  \\
{[\ion{O}{3}]}$\lambda5007$     & $\Gemfluxfivezerozeroseven{}$ \\
{[N II]}$\lambda6548$     & $\Gemfluxsixfivefoureight{}$  \\
H$\alpha$          & $\Gemfluxsixfivesixtwo{}$   \\
{[N II]}$\lambda6584$     & $\Gemfluxsixfiveeightfour{}$  \\
{[S II]}$\lambda6717$     & $\Gemfluxsixsevenoneseven{}$  \\
{[S II]}$\lambda6731$     & $\Gemfluxsixseventhreeone{}$  
\enddata
\tablecomments{All fluxes have been measured from our Gemini spectrum. No reddening correction has been applied. There is an additional systematic uncertainty in the flux measurements of $\sim10$\% due to flux calibration.}

\end{deluxetable}

\subsection{Radio Observations} \label{obs:rad}

We observed SN\,2010ay with the EVLA \citep{EVLA} on three epochs, 2010
March 26, 2010 April 29, and 2011 May 7. All EVLA observations were
obtained with a bandwidth of 256 MHz centered at 4.9 GHz. We used
calibrator J1221+2813 to monitor the phase and 3C286 for flux
calibration. Data were reduced using the standard packages of the
Astronomical Image Processing System (AIPS).  We do not detect a radio
counterpart to SN\,2010ay in these observations and place upper limits
of $F_\nu \lesssim 46, 42, 30~\mu$Jy ($3\sigma$) for each epoch
respectively corresponding to upper limits on the spectral luminosity
spanning $L_{\nu}\lesssim (3.6-5.5)\times 10^{27}~\rm
erg~cm^{-2}~s^{-1}$.

\begin{figure*}
\epsscale{.75}
\plotone{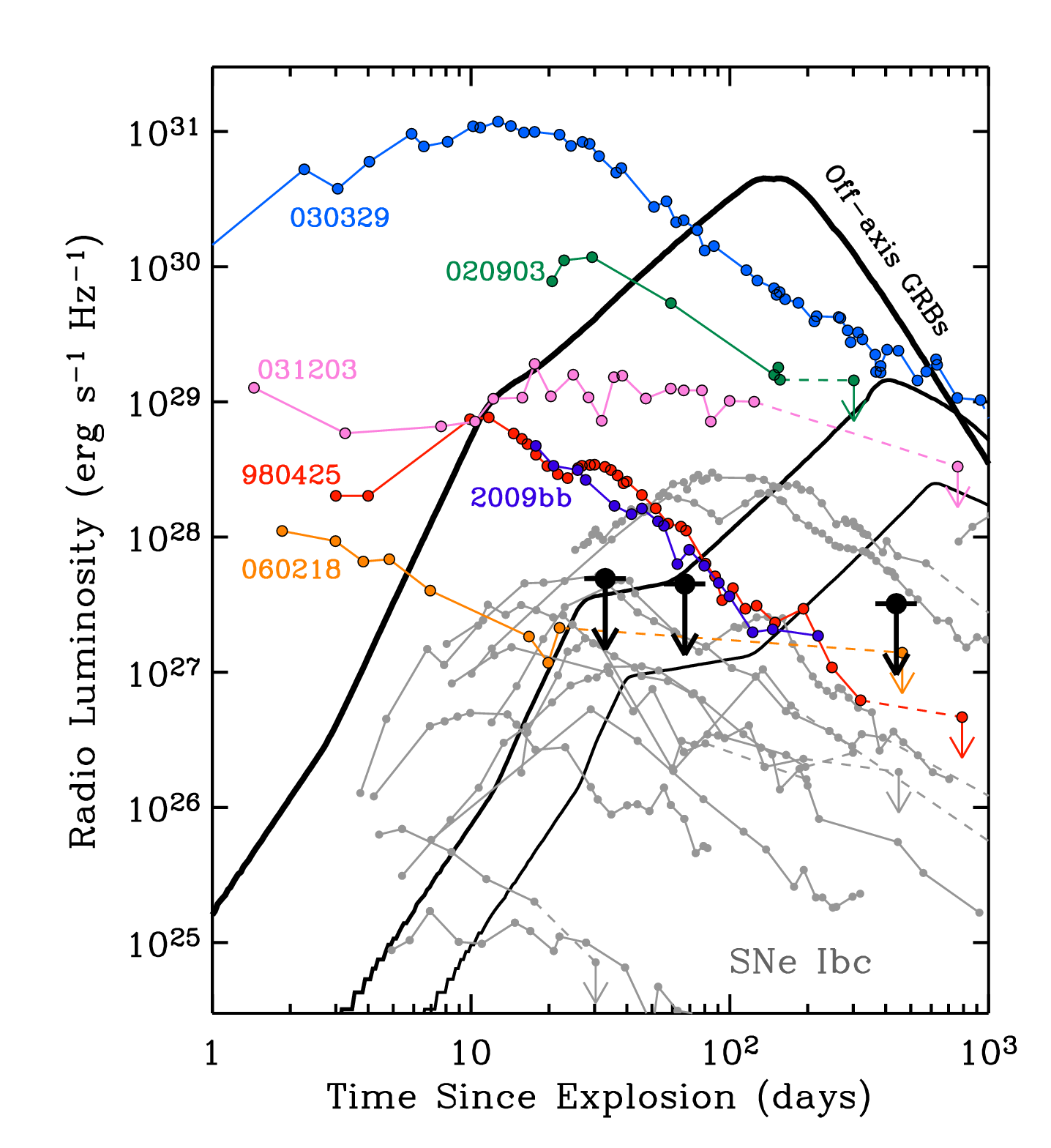}
\caption{EVLA upper limits for SN\,2010ay (black arrows) are compared with off-axis GRB afterglow lightcurve models (black curves; $30,~60,~90^\circ$) and the observed radio light-curves for ordinary SNe Ib/c (grey; \citealt{ams07} and references within) and the radio afterglows of all GRB-SNe within $z\le 0.25$. SN\,2010ay is a factor of $10^2$ to $10^3$ less luminous than XRF\,020903 (orange; \citealt{skb+04a}), GRB\,030329 (blue; \citealt{bkp+03,fsk+05}), and GRB\,031203 (\citealt{skb+04b}). Relativistic, engine-driven SNe 1998bw (red; \citealt{kfw+98}) and 2009bb (dark blue; \citealt{scp+10}) are a factor of 10 more luminous than the SN\,2010ay limits on a comparable timescale, while XRF\,060218 lies a factor of a few below the limits. We constrain the radio counterpart to be no more luminous than XRF\,060218 and comparable to the peak luminosities of ordinary SNe Ib/c.\label{fig:radio1}}
\end{figure*}

As shown in Figure~\ref{fig:radio1}, these limits are comparable to
the peak luminosities observed for ordinary SNe Ib/c
(\citealt{bkf+03,ams07,cf06,scp+10} and references within) and a factor of
$10^2$ to $10^3$ less luminous than the radio afterglows associated
with GRBs 020903, 030329, and 031203 at early epochs
\citep{bkp+03,skb+04a,skb+04b,fsk+05}. In comparison with the
radio luminosities observed for the relativistic SNe 1998bw
\citep{kfw+98} and 2009bb \citep{scp+10}, SN\,2010ay is a factor of 
$\gtrsim 10$ less luminous. The peak radio luminosity observed for GRB 100316D was a factor of a few higher than the first EVLA non-detection of SN\,2010ay \citep{Soderberg10bh}. The only relativistic explosion with
detected radio emission below our EVLA limits is the weak and fast
fading XRF\,060218 \citep{skn+06}.

\section{Initial Constraints} \label{res:snprop}

\subsection{Light Curve Modeling} \label{exp:lcmodel}

We construct an R-band lightcurve for SN\,2010ay using the observations
described in \S\ref{obs:opt} (Table \ref{tab:phot}). We convert the
\ips\ and $r^\prime$ band data points to the $R-$band using the fiducial lightcurve method of \cite{Soderberg040924}, assuming the unextincted $i^\prime-R$ and $r^\prime-R$ colors
observed for the Type Ic-BL SN\,1998bw at the appropriate epoch
\citep{Galama98}. Photometry for the unfiltered CRTS images was reported by \cite{DrakeCBET} after transformation to the synthetic V-band (A.J. Drake, private communication); we therefore convert to the $R-$band assuming the $V-R$ color of SN 1998bw at the appropriate epoch.  For the \PS\ photometry, the host galaxy flux was subtracted using a template
image. For all other photometry, we have
subtracted the flux of the host galaxy numerically assuming the
magnitude reported in Table \ref{tab:SDSS}. A total (Galactic + host)
reddening of E(B-V)$=\GemEBmV$~mag has been assumed (see
\S\ref{obs:host}).

To estimate the explosion date of SN\,2010ay we have fit an expanding
fireball model to the optical light curve (Figure \ref{fig:lc}),
following \cite{Conley06}. In this model, the luminosity increases
as 

\begin{equation}
L\propto \Bigg(\frac{t-t_0}{1+z}\Bigg)^n
\end{equation}

\noindent We derive an explosion date $t_0$ of \texp{}. Here we have assumed an index $n=2$. This suggests that the PS1 $3\pi$ \ips\-band pre-discovery
detection image of SN\,2010ay was taken $\sim\texpthreepii{}$ days
after the explosion. Observations by the PS1 survey have therefore provided a valuable
datapoint for estimating the explosion date and also for constraining
the rise-time of SN\,2010ay, as well as other nearby SNe \citep[e.g. SN\,2011bm][]{Valenti12}.
 
In Figure~\ref{fig:lc}, we compare the light curve of SN\,2010ay to the
SN Ib/c light-curve template of \citet{Drout10} after stretching by a
factor of $(1+z)$. The template provides a reasonable fit to the optical
evolution of SN\,2010ay. Fitting the template to our photometry we
derive (reduced $\chi^2=\lcchi{}$) a date of R-band peak of 2010 March
$18\pm 2$ UT (MJD $55273\pm 2$) and a R-band peak magnitude of
$M_R\approx -19.7$~mag before extinction correction. As discussed in
\S\ref{obs:host}, based on the Balmer decrement observed for the host
galaxy emission lines, we assume an extinction of
$E(B-V)=\GemEBmV{}$~mag. Applying this correction, the peak absolute
magnitude is $M_R\approx\Rpeak{}$~mag.  We note that this fitted value is
$\approx 0.2$ mag fainter than that estimated from the data point
near peak. Here, the uncertainty is dominated by the template fitting.

Regardless of the extinction correction, SN\,2010ay is more luminous than
all the 25 SNe Ib/c in the sample of \cite{Drout10}, except for SN
2007D ($M_R\approx -20.65$~mag, which was also significantly extincted:
$A_V\sim1.0$~mag). Assuming an intrinsic V-R color of zero at peak
\citep[e.g. 1998bw:][]{Galama98,Patat01}, SN\,2010ay is also more luminous
than any of the 22 GRB and XRF-producing SNe in the compilation of
\cite{Cano112}, all corrected for extinction, and is \Canosdev{}
standard deviations from the mean luminosity. The peak magnitude is
only $\sim 1$ mag below that of the Type Ic SN\,2007bi
($M_R=-21.3\pm0.1$~mag), which \cite{GalYam09} report as
a candidate pair-instability supernova.

\subsection{Large Nickel Mass for SN\,2010ay} \label{res:explosion}

We use the available photometry for SN\,2010ay discussed above to
derive the mass of $^{56}$Ni required to power the optical light-curve
under the assumption that the emission is powered by radioactive
decay.  Using the relation between $M_{\rm Ni}$ and $M_{R}$ found by
\citet{Drout10}, $\log(M_{\rm Ni})=(-0.41~M_{R}-8.3)~M_{\odot}$, we estimate
that SN\,2010ay synthesized a nickel mass of $M_{\rm
Ni}=\Mni{}~M_{\odot}$.  We have estimated the uncertainty in the
$M_{\rm Ni}$ estimate by propagation of the uncertainty in the template fitted
peak magnitude --- systematic uncertanties are not included.   If we instead adopt the most luminous individual data point in the light curve as the peak value, rather than the smaller peak value from template fitting, we find $M_{\rm Ni}\approx1.2~M_{\odot}$.

The $M_{\rm Ni}$ estimate for SN\,2010ay is larger than that of all
but one (SN\,2007D) of the 25 SNe Ib/c in the \cite{Drout10} compilation
and significantly larger than the estimate for GRB-SN\,2010bh, $M_{\rm
Ni}=0.12\pm0.01 M_{\odot}$ \citep{Cano11}. On the other hand, $M_{\rm Ni}$ of SN\,2010ay is at least $3\times$ smaller than for SN\,2007bi
\cite[$M_{\rm Ni}\approx3.5-4.5~M_\odot$,][]{Young10}. A pair
instability supernova should produce $M_{\rm Ni}\gtrsim3~M_\odot$
\citep{GalYam09}. 

\subsection{Unusually high velocity} \label{obs:vel}

As illustrated in Figure \ref{fig:spectra}, the broad, highly-blended
spectral features of SN\,2010ay at the time of the WHT observations
(\tpeakWHT{} days after R-band peak, see \S\ref{exp:lcmodel}) resemble
those of SN\,2010bh at a similar time \citep[10.0 d after
peak,][]{Chornock2010bh}. In particular, the blueshift
of the feature near 6355~$\rm \AA$ is larger than in
SN 1998bw and more similar to SN\,2010bh. This feature is commonly associated with
\ion{Si}{2}~$\lambda6355\rm \AA$ \citep[e.g.][]{Patat01}. However, this
feature has two clearly-detectable absorption minima in SN\,2010ay, but
not in SN\,2010bh. This could be due to increased blending in SN\,2010bh
or the absence of contaminating lines. The red ends of the SN\,2010ay
and SN\,2010bh spectra (rest wavelength $>7500~\rm \AA$) have similar
P-Cygni features, but the emission and absorption components in SN
2010bh are each blueshifted by $\sim200~\rm \AA$ more than in the spectrum
of SN\,2010ay. \cite{Chornock2010bh} attribute this feature to the
\ion{Ca}{2} NIR triplet, with a $gf$-weighted line centroid of
$8479~\rm \AA$, and find a velocity that is high, but consistent with the
early-time velocity of \ion{Si}{2}~$\lambda6355\rm \AA$
($30-35\times~10^3$~km~s$^{-1}$).

We measure the absorption velocity from the minimum of the \ion{Si}{2}~$\lambda6355\rm \AA$ absorption feature ($\vSi$). We smooth the spectrum using an inverse-variance-weighted Gaussian filter \citep[][with $d\lambda/\lambda=0.005$]{Blondin06} and measure the minimum position of the redmost component of the absorption profile. The blue component of the absorption profile shifts blueward over time, suggesting that it is produced by a combination of ions separated by several $10^3$~km~s$^{-1}$, such as \ion{He}{1}~$\lambda5876\rm \AA$ and \ion{Na}{1} D, whose relative optical depth changes with time.

The absorption velocity inferred from the
\ion{Si}{2}~$\lambda6355\rm \AA$ feature is $\sim2\times$ faster than
that of SN 1998bw at similar times, and more similar to that of SN
2010bh (Figure~3). For SN 1998bw, \cite{Patat01} measured
$\sim10\times10^3$~km~s$^{-1}$ at +13 days. For SN\,2010bh,
\cite{Chornock2010bh} measured $\vSi\approx
26\times10^3$~km~s$^{-1}$ at +10.0 days after
explosion. \cite{PrietoCBET} reported a velocity of
$\vSi\approx$22.6$\times10^3$~km~s$^{-1}$ from the
\ion{Si}{2}~$\lambda6355\rm \AA$ feature in a spectrum of SN\,2010ay taken
at $+\tpeakPrieto{}$~days; \cite{PrietoCBET} do not discuss the details of their methodology for the velocity measurement. From our [WHT,Gemeni] spectra taken
$[+\tpeakWHT{},+\tpeakGemini{}]$ days after R-band peak (see
\S\ref{exp:lcmodel}), we estimate
$\vSi\approx$[\VelWhtTonrySiA,\VelGemTonrySiA]$\times10^3$~km~s$^{-1}$.

In addition to the broadening of the spectral features and the blueshift of the \ion{Si}{2}~$\lambda6355\rm \AA$ line, additional lines of evidence suggest a high photospheric expansion velocity for SN\,2010ay. We measure $\vSi\approx[\VelWhtTonryCa,\VelGemTonryCa]\times10^3$~km~s$^{-1}$ from the \ion{Ca}{2} NIR triplet on the smoothed [WHT,Gemini] spectra, relative to a line center at $8479~\rm \AA$. This is within a few $10^3$~km~s$^{-1}$ of the $\vSi$ we measure from \ion{Si}{2}~$\lambda6355\rm \AA$ at these epochs. Furthermore, we do not detect the broad emission ``bump'' near $4500~\rm \AA$ in either of our spectra. This feature was also absent in SN\,2010bh, but was identified in SN\,2003dh, SN2006aj, and several Ic-BLs not associated with GRBs and normal SNe~Ic;  \cite{Chornock2010bh} suggest that the absence of this feature indicates a high expansion velocity if it is due to blending of the iron lines to the blue and red of $4500~\rm \AA$.

\begin{figure*}
\plotone{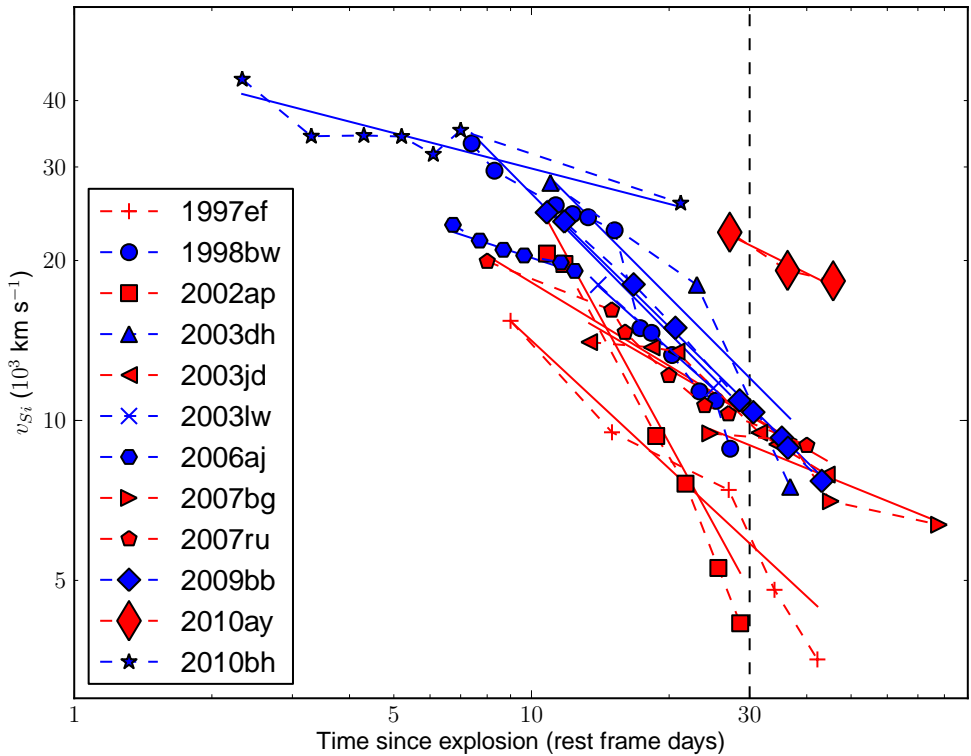}
\caption{A comparison of the time-evolving absorption velocity of SN\,2010ay and other SNe~Ic-BL (red) and engine-driven explosions (blue) from the literature. For each SN, we fit a power law of the form $\vSi=\vSi^{30}(t/30)^{\alpha}$, where t is the time in days, $\vSi^{30}$ is the velocity at 30 days after explosion (dashed vertical line), and $\alpha$ is the velocity gradient. The velocities for SNe 1997ef, 2003dh, 2003lw are from \cite{Mazzali06}, as determined by spectral modeling. The velocities for all other SNe are measured from the \ion{Si}{2}~$\lambda6355\rm \AA$ feature as follows: SN\,2007ru are from \cite{Sahu09}; SNe 1998bw, 2006aj, and 2010bh are from \cite{Chornock2010bh}, from spectra in references therein; SN\,2007bg are from \cite{Young10}; SNe 2002ap, 2009bb, and 2003jd are from \cite{Pignata11} and references therein; SN 2010ay is from \cite{PrietoCBET} and this paper.  See Table \ref{tab:vthirty} for uncertainties.}\label{fig:vc}
\end{figure*}

We compare the late-time expansion velocity of SN\,2010ay to other SNe
Ic-BL and GRB-SNe with detailed, multi-epoch velocity measurements
from the literature in Figure \ref{fig:vc}. 
We note that the velocities for SNe 1997ef, 2003dh, 2003lw plotted in the figure were estimated by \cite{Mazzali06} via spectral modeling, rather than measured directly from the minimum of the \ion{Si}{2}~$\lambda6355\rm \AA$ absorption feature; however, these velocities should be equivalent to within a few $10^3$~km~s$^{-1}$, as the minimum of the \ion{Si}{2} feature is typically well-fit by the photospheric velocity of the spectral models \citep[see e.g.][]{Mazzali00,Kinugasa02}.  
In this figure, we also
fit power-law gradients to the time-evolution of the velocity of these
SNe with the form $\vSi=\vSi^{30}(t/30)^{\alpha}$, where t is the
time since explosion in days and $\vSi^{30}$ is the velocity at 30
days in units of $10^3$~km~s$^{-1}$.  These parameters are listed in Table \ref{tab:vthirty}.  Figure \ref{fig:vc} illustrates
that most SNe are well described by a single power law.\footnote{A
break at $\sim2\times10^4$~km~s$^{-1}$ appears to exist for SN 1998bw
at $\sim16$~days, as noted by \cite{Kinugasa02}.} However, due to lack of late-time spectroscopy, the $\vSi^{30}$ measurement amounts to an extrapolation for some objects (particularly 2006aj), and contamination from different ions or detached
features will add uncertainty to velocities measured from the
\ion{Si}{2}~$\lambda6355\rm \AA$ feature.

\begin{deluxetable}{p{90pt}rr}
\tablecaption{Velocity Evolution of SNe~Ic-BL\label{tab:vthirty}}
\tablewidth{0pt}
\tabletypesize{\scriptsize}
\tablehead{ \colhead{SN} & \colhead{$\vSi^{30}$} & \colhead{$\alpha$}}
\startdata
\cutinhead{SNe~Ic-BL}
1997ef  &   $\vthirtyef$    &   $\vslopeef$    \\
2002ap  &   $\vthirtyap$    &   $\vslopeap$    \\
2003jd  &   $\vthirtyjd$    &   $\vslopejd$    \\
2007bg  &   $\vthirtybg$    &   $\vslopebg$    \\
2007ru  &   $\vthirtyru$    &   $\vsloperu$    \\
2010ay  &   $\vthirtyay$    &   $\vslopeay$    \\
\cutinhead{Engine-driven SNe~Ic-BL}
1998bw  &   $\vthirtybw$    &   $\vslopebw$    \\
2003dh  &   $\vthirtydh$    &   $\vslopedh$    \\
2003lw  &   $\vthirtylw$    &   $\vslopelw$    \\
2006aj  &   $\vthirtyaj$    &   $\vslopeaj$    \\
2009bb  &   $\vthirtybb$    &   $\vslopebb$    \\
2010bh  &   $\vthirtybh$    &   $\vslopebh$    \\
\enddata
\tablecomments{To the velocity measurements for each SN, we have fit a power law of the form $\vSi=\vSi^{30}(t/30)^{\alpha}$, where t is the time since explosion in days and $\vSi^{30}$ is the velocity at 30 days in units of $10^3$~km~s$^{-1}$.  These power law fits are illustrated in Figure \ref{fig:vc}.  We have assumed $10\%$ uncertainties in all velocity measurements.  Uncertainties in the explosion date for SNe without detected GRBs vary due to availability of early-time photometry; we adopt the following conservative uncertainties: 7~d (1997ef, \citealt{Hu97}), 7~d (2002ap, \citealt{GalYam02}), 6~d (2003jd, \citealt{Valenti08}), 14~d (2007bg), 3~d (2007ru, \citealt{Sahu09}), 1~d (2009bb, \citealt{Pignata11}), 2~d (2010ay; this paper).}

\end{deluxetable}

SN\,2010ay and 2010bh share high characteristic velocities at 30 days after
explosion and velocity gradients that are low relative
to other broad-lined Ic SNe with and without associated GRBs. For SN
2010ay, $\vSi^{30}=\vthirtyay$ is $2-4\times$ larger than for other
SNe~Ic-BL without associated GRBs ($\vSi^{30}=\vthirtyef$ for 1997ef,
$\vthirtyap$ for 2002ap, $\vthirtyjd$ for 2003jd, $\vthirtybg$ for
2007bg, and $\vthirtyru$ for 2007ru) and is similar to the GRB-SN
2010bh ($\vSi^{30}=\vthirtybh$). No other GRB-SN or SN~Ic-BL has
$\vSi^{30}>15$. The SNe~Ic-BL and GRB-SNe with the most shallow
velocity gradients among these twelve objects have $\alpha<-0.5$; they
are SN\,2006aj ($\alpha=\vslopeaj$), SN\,2007bg ($\alpha=\vslopebg$), SN
2010ay ($\alpha=\vslopeay$), and SN\,2010bh ($\alpha=\vslopebh$). This places SN\,2010ay in the company of two GRB-SNe in having a slowly-evolving absorption velocity, and SN\,2007bg (whose unusually-fast decline rate distinguishes it from other SNe~Ic-BL, \citealt{Young10}). The velocities of SNe 2010ay and 2010bh, respectively,
decline about 2 and $4\times$ more slowly than the other SNe~Ic-BL
(mean and standard deviation: $\alpha=\velfitslopeicblmean \pm
\velfitslopeicblstd$) and about 1.5 and $3\times$ more slowly the other
GRB-SNe ($\alpha=\velfitslopegrbmean \pm
\velfitslopegrbstd$).  The slow \ion{Si}{2}~$\lambda6355\rm \AA$ absorption velocity evolution of SN\,2010ay at late times resembles the slow evolution of the \ion{Fe}{2} lines of the spectroscopically-normal Type Ic SNe\,2007gr and 2011bm at late times \citep{Valenti12}.

Given the high peak luminosity of SN\,2010ay (\S\ref{exp:lcmodel}), we also consider the velocity of the candidate pair-instability SN\,2007bi.  Velocity measurements for SN\,2007bi are only
available at late times \citep[$>50$ d,][]{Young10}. Fitting to these
late-time \ion{Si}{2}~$\lambda6355\rm \AA$ velocity measurements, we find
that SN\,2007bi has a characteristic velocity $\sim3\times$ smaller
than 2010ay: $\vSi^{30}=\vthirtybi$ and the late-time velocity gradient
is $\sim2\times$ more shallow: $\alpha=\vslopebi$.

\subsection{Ejecta Mass and Energy}

We use the scaling relations provided by
\cite{Drout10}, based on the original formalism of \cite{Arnett82} and
modified by \cite{Valenti08}, to derive the total mass of the ejecta
and the kinetic energy.

\begin{eqnarray}
M_{\rm ej}=0.8\Bigg(\frac{\tau_c}{8 \rm d}\Bigg)^2\Bigg(\frac{\vSi}{\rm 10,000~km~s^{-1}}\Bigg)~M_{\odot}\\
E_{K}=0.5\Bigg(\frac{\tau_c}{8 \rm d}\Bigg)^2\Bigg(\frac{\vSi}{\rm 10,000~km~s^{-1}}\Bigg)^3~\times10^{51}~{\rm~erg}
\end{eqnarray}

We assume the fitted peak magnitude for SN\,2010ay ($M_R=\Rpeak$, \S\ref{exp:lcmodel}), the absorption velocity we measure from our WHT spectrum at $\tpeakWHT$~days after maximum light ($\vSi=\VelWhtTonrySiA{}\times10^3$~km~s$^{-1}$, \S\ref{obs:vel}), and a characteristic time (light-curve width) consistent with the data and the mean value from the \cite{Drout10} sample of SNe~Ic-BL ($\tau_c=14$ d).

Using these values, the total mass ejected was $M_{\rm ej}\approx \Mej{}~M_{\odot}$, and the total kinetic energy of the explosion was $E_{K}\approx \Ek{}\times10^{51}$~ergs.  Hereafter we refer to the definition $E_{K,51}=E_{K}/10^{51}$~ergs.

The systematic uncertainties associated with this modeling dominate
the statistical uncertainties. In particular, the models rely on the
assumptions of homologous expansion, spherical symmetry, all $^{56}$Ni
centralized at the center of the ejecta, optically thick ejecta and
constant opacity.

We note that an earlier measurement of the absorption velocity is
preferred for optical modeling.  Since we have argued that SN
2010ay and SN\,2010bh have similar characteristic velocities
($\vSi^{30}$), if we instead adopt a higher velocity of
$25,000$~km~s$^{-1}$ as measured for SN\,2010bh by \cite{Cano11}, we
estimate $M_{\rm ej}\approx\Mejbh{}~M_{\odot}$ and
$E_{K,51}\approx\Ekbh{}$ for SN\,2010ay.

The $M_{\rm ej}$ and $E_{K,51}$ of SN\,2010ay are consistent
with the mean for SNe~Ic-BL in the \cite{Drout10} sample ($4.7^{+2.3}_{-1.8}~M_\odot$
and $11^{+6}_{-4}$, respectively), because the authors assumed a
velocity ($\vSi=2\times10^4$~km~s$^{-1}$) similar to the late-time
velocity we measure. For comparison, SN\,2010bh had a total ejecta
mass of $\sim2~M_\odot$ and a total kinetic energy of
$E_{K,51}\approx13$ \citep{Cano11}.

The ratio of Ni to total ejecta mass is $\sim \niejratio{}$ for SN
2010ay, significantly higher than the values typical of SNe~Ic-BL and
GRB-SNe.  For comparison, the ratio is just $\sim0.05$ for SN
2010bh. Adopting the values derived from bolometric light curve
modeling by \cite{Cano11}, the $M_{\rm Ni}$ and $M_{\rm Ni}/M_{\rm
ej}$ ratios for other GRB-SNe are: $\sim0.5~M_\odot$ and
$\sim0.06-0.22$ (1998bw), $\sim0.4~M_\odot$ and $\sim0.08$ (2003dh),
$\sim0.15~M_\odot$ and $\sim0.07-0.1$ (2006aj), and $\sim0.2~M_\odot$
and $\sim0.06$ (2009bb). This ratio for SN\,2010ay is larger than that
of all but 4 of the 25 SNe of
\cite{Drout10}: the Type Ic SNe 2004ge ($M_{\rm Ni}/M_{\rm
ej}\sim0.4$) ans 2005eo ($M_{\rm Ni}/M_{\rm ej}\sim0.2$), the Type Ib
SN\,2005hg ($\sim0.4$), and the Type Ic-BL SN\,2007D ($\sim0.6$).

The large value of $M_{\rm Ni}$ we estimate for SN\,2010ay raises the
question of whether a process other than Ni decay may be powering its
light-curve \citep[e.g.][]{Chatzopoulos09}. An independent test of the physical process powering the
light-curve is the decay rate of the late-time light curve which
should be $\approx 0.01$~mag~day$^{-1}$ for SNe powered by radioactive
decay of $^{56}$Co. For SN\,2007bi, \cite{GalYam09} derive $M_{\rm
Ni}=3.5~M_\odot$ from the measured peak magnitude and find that the
late-time light curve is consistent with the decay rate of
$^{56}$Co. While the \PS\ $3\pi$ survey also observed the field in
2011 March, the SN was not detected in our subtracted images and
the limits are not constraining in the context of $^{56}$Co decay (see
\S\ref{obs:photps1} and Table \ref{tab:phot}). Another possible process is a radiation-dominated
shock that emerges due to interaction with an opaque circumstellar
medium, as has recently been proposed by \cite{Chevalier11} for the
ultra-luminous SNe 2006gy and 2010gx
\citep{Pastorello10,Quimby11}. However, while this class of
ultra-luminous objects shares some spectroscopic similarities to SNe~Ic
\citep{Pastorello10}, they show peak luminosities
$\sim4-100\times$ higher than SN\,2010ay \citep{ccs+11,Quimby11}.

\section{Constraints on Relativistic Ejecta} \label{res:evla}

We use our EVLA upper limits for SN\,2010ay spanning $\Delta t\approx
29-433$ days to constrain the properties of the shockwave and those of
the local circumstellar environment. The radio emission from SNe Ib/c
and GRBs is produced by the dynamical interaction of the fastest
ejecta with the surrounding material \citep{c82}. The kinetic energy of the ejecta is converted, in part, to
internal energy of the shocked material which itself is partitioned
between relativistic electrons ($\epsilon_e$) and amplified magnetic
fields ($\epsilon_B$). Following the
breakout of the shockwave from the stellar surface, electrons in the
environment of the explosion are shock-accelerated to relativistic
velocities with Lorentz factor, $\gamma_e$ and distributed in a
power-law distribution characterized by $N(\gamma_e)\propto
\gamma_e^{-p}$. Here, $p$ characterizes the electron energy index. The
particles gyrate in amplified magnetic fields and give rise to
non-thermal synchrotron emission that peaks in the radio and mm-bands
in the days to weeks following explosion with observed spectral index,
$F_{\nu}\propto
\nu^{-(p-1)/2}$. At lower frequencies the emission is suppressed due
to synchrotron self-absorption which defines a spectral peak, $\nu_p$
\citep{c98}. 

The dynamics of the shockwave determine the evolution of the
synchrotron spectrum, and in turn, the properties of the observed
radio light-curves. In the case of SNe Ib/c, there are three primary
scenarios for the dynamical regime of the ejecta depending on the
shock velocity, $v=\beta c$ (associated Lorentz factor, $\Gamma$): (i)
non-relativistic ($v\approx 0.2c$) free-expansion as in the case of
ordinary SNe Ib/c \citep{c98}, (ii) a decoupled and relativistic
($\Gamma\sim 10$) shell of ejecta that evolves according to the
Blandford-McKee solution for several months
\citep{spn98} before transitioning to the Sedov-Taylor regime
\citep{fwk00}. This is the standard scenario for typical GRBs. And
(iii) a sub-energetic GRB with trans-relativistic velocity
($\beta\Gamma\lesssim 3$) that bridges the free-expansion and
Blandford McKee dynamical regimes (e.g.\,SN\,1998bw;
\citealt{kfw+98}, \citealt{lc99}).

We consider our EVLA upper limits in the context of these three models
below. For shock velocities of $v\gtrsim 0.2c$, $\epsilon_e\approx
0.1$ is reasonable
\citep{skb+05,cf06}. 
We further assume equipartition, $\epsilon_e=\epsilon_B=0.1$. We adopt
a free expansion model for both the non-relativistic ordinary SN Ib/c
case and the sub-energetic, trans-relativistic GRB scenario. As shown
by \cite{lc99} a free-expansion model is still reasonable in the
trans-relativistic regime (cases iii, see above).

\subsection{Freely-expanding shockwave}

In the free-expansion scenario, a shock discontinuity separates the
forward and reverse shocks, located at the outer edge of the stellar
envelope. The bulk ejecta is in free expansion while the thin layer
of post shock material is slightly decelerated, $R\propto t^{0.9}$
\citep{cf06}. At a given frequency, the bell-shaped light-curves of
the SN synchrotron emission may be described as \citep{c98}

\begin{equation}
L_{\nu}\approx 1.582\times L_{\nu,p} \left(\frac{\Delta t}{t_p}\right)^a \left[1-e^{-(\Delta t/t_p)^{-(a+b)}}\right]
\label{eqn:lc}
\end{equation}

\noindent
where $L_{\nu,p}$ is the flux density at the spectral peak at epoch,
$t_p$.  Assuming an electron index of $p\approx 3$, consistent with
radio spectra of SNe Ib/c \citep{cf06}, the exponents are $a\approx
2.3$ and $b\approx 1.3$. The time averaged shockwave velocity is
$\overline{v}\approx R/\Delta t$ where $R$ is the shockwave radius
defined as

\begin{eqnarray}
R&\approx &2.9\times 10^{16}\left(\frac{\epsilon_e}{\epsilon_B}\right)^{-1/19} \left(\frac{L_{\nu,p}}{10^{28}~\rm~erg~s^{-1}~Hz^{-1}}\right)^{9/19}\nonumber\\
&&\times\left(\frac{\nu_p}{5~{\rm GHz}}\right)^{-1}~\rm cm.
\label{eqn:radius}
\end{eqnarray}

Here we make the assumption that the radio emitting region is half of
the total volume enclosed by a spherical blastwave. Next, we estimate
the internal energy, $E$, of the radio emitting material from the
post-shock magnetic energy density, $E\approx B^2 R^3/12\epsilon_B$
where we maintain the assumption of equipartition. As shown by
\citet{c98}, the amplified magnetic field at peak luminosity is also directly determined from
the observed radio properties,

\begin{eqnarray}
B&\approx& 0.43~\left(\frac{\epsilon_e}{\epsilon_B}\right)^{-4/19} \left(\frac{L_{\nu,p}}{10^{28}~{\rm~erg~s^{-1}~Hz^{-1}}}\right)^{-2/19} \nonumber\\
&&\times\left(\frac{\nu_p}{5~{\rm GHz}}\right)~\rm G. 
\label{eqn:Bfield}
\end{eqnarray}

Finally, the mass loss rate of the progenitor star, $\dot{M}$, may be
derived from the number density of emitting electrons. Here we
normalize the wind profile according to $\rho\propto Ar^{-2}$ and
$A_*=A/5\times 10^{11}~\rm g~cm^{-}$. This normalization of $A_*$ implies that an $A_*$ of $1$ corresponds to typical Wolf-Rayet progenitor wind properties of $\dot{M}=10^{-5}~\rm M_{\odot}~yr^{-1}$ and a progenitor wind velocity of $v_w=10^3~\rm km~s^{-1}$.

\begin{eqnarray}
A_* &\approx& 0.15~\left(\frac{\epsilon_B}{0.1}\right)^{-1} \left(\frac{\epsilon_e}{\epsilon_B}\right)^{-8/19} \left(\frac{L_{\nu,p}}{10^{28}~{\rm~erg~s^{-1}~Hz^{-1}}}\right)^{-4/19} \nonumber\\
&&\times\left(\frac{\nu_p}{5~{\rm GHz}}\right)^2 \left(\frac{\Delta t}{10~\rm days}\right)^2 ~{\rm cm}^{-1} 
\end{eqnarray}

\noindent where we assume a shock compression factor of $\sim 4$ and 
an nucleon-to-electron ratio of two.

We built a two-dimensional grid of fiducial radio light-curves
according to Eqn.~\ref{eqn:lc} in which we vary the parameters
$L_{\nu_p}$ and $\nu_p$ over a reasonable range of parameter space,
bounded by $t_p\approx [1,3000]$ days and $F_{\nu,p}\approx 0.04-1000$
mJy. We identify the fiducial light-curves associated with a radio
luminosity {\it higher} than the EVLA upper limits for SN\,2010ay at
each epoch as these are excluded by our observations. We extract the
physical parameters associated with these excluded light-curves ($R$,
$B$, $E$, $A_*$) to define the parameter space excluded by our radio
observations. The parameter space for $\nu_p$ and $F_{\nu,p}$ are
bounded by the respective values for which the model exceeds
relativistic velocities, $\beta\Gamma\sim$ few.

\begin{figure}
\plotone{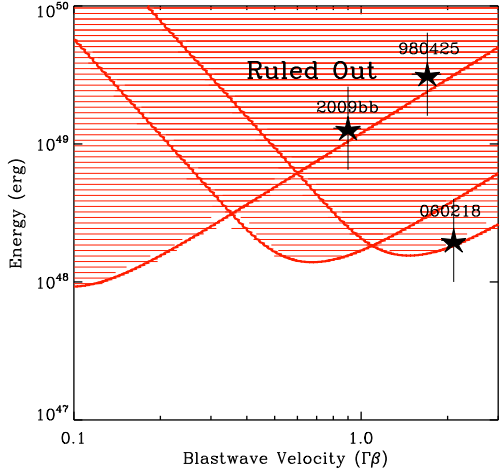}
\caption{The region of energy-velocity space ruled out (red) by our EVLA observations for on-axis ejecta under the assumption of a free-expansion model.\label{fig11}}
\label{fig:ev_sn2010ay}
\end{figure}

As shown in Figure~\ref{fig:ev_sn2010ay}, our deep EVLA limits enable
us to rule out a scenario in which there is copious energy, $E\gtrsim
10^{48}$~erg, coupled to a relativistic outflow, in this
two-dimensional $E-v$ parameter space. The excluded region includes
GRB-SNe 1998bw and 060218 as well as the relativistic SN\,2009bb. It
does not exclude the standard scenario in which a small percent of the
energy is coupled to fast moving material within the homologous
outflow, as is typically observed for ordinary SNe Ib/c ($E\approx
10^{47}$ and $v\approx 0.2c$; \citealt{scp+10}).

\begin{figure}
\plotone{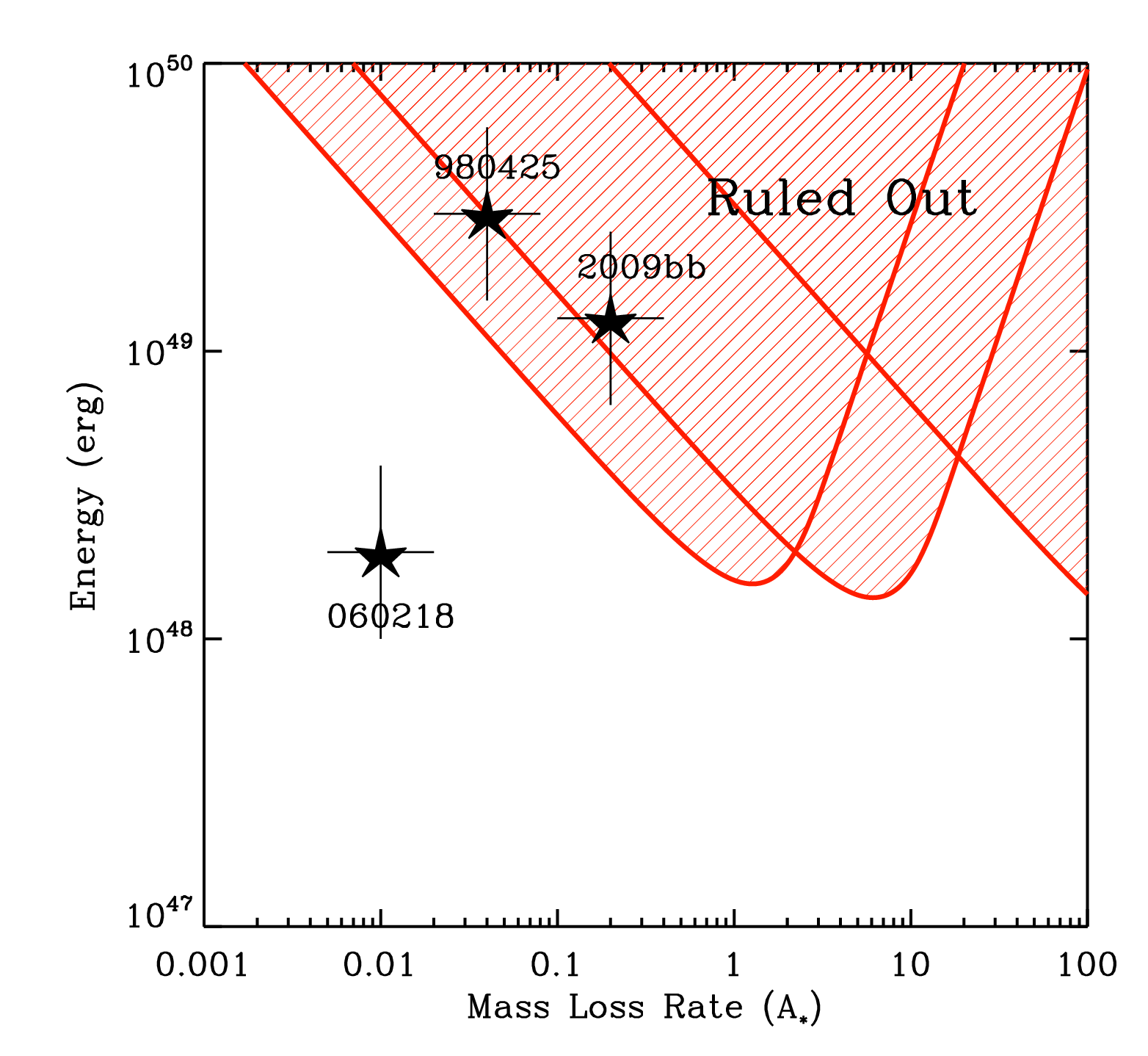}
\caption{The region of energy-mass loss space ruled out (red) by our EVLA observations for on-axis ejecta under the assumption of a free-expansion model.\label{fig10}}
\label{fig:ed_sn2010ay}
\end{figure}

Next we consider the effects of circumstellar density since lower mass
loss rates produce fainter radio counterparts. As shown in
Figure~\ref{fig:ed_sn2010ay}, the EVLA limits for SN\,2010ay exclude
the region of parameter space populated by SNe 1998bw and 2009bb with
mass loss rates of $A_*\sim 0.1$,
however, the low density environment of GRB\,060218 lies outside of
our excluded region due to its lower CSM density,
$A_*\sim 0.01$ that gives rise to a lower
luminosity radio counterpart.

\subsection{Relativistic Ejecta}
\label{sec:offaxis}

In the case of relativistic deceleration the ejecta are confined
to a thin jet and are physically separated from the homologous SN
component. Deceleration of the jet occurs on a timescale of $\Delta
t\approx (E_{51}/A_*)$ years in a wind-stratified medium
\citep{wax04}. On this same timescale, any ejecta components 
that were originally off-axis spread sideways into the observer's
line-of-sight. While the early EVLA limits constrain the properties
of the on-axis ejecta according to the free-expansion model described
above, the late time EVLA upper limits constrain any radio emission
from a GRB jet originally pointed away from our line-of-sight.

For this scenario, we adopt the semi-analytic model
of \citet{snb+06} for off-axis GRB jets.
In Figure~\ref{fig:radio1} we compare the radio upper limits for SN\,2010ay
with the predictions for an off axis jet with standard parameters
($E_K\approx 10^{51}$ erg, $A_*=1$, $\epsilon_e=\epsilon_B=0.1$, $\theta_j=5^o$)
and assuming a viewing angle of $\theta_{\rm oa}=30, 60,$ or 90 degrees.
Our EVLA upper limits rule out all three of these model light-curves.  
We derive the two-dimensional parameter space (energy and
CSM density) that is excluded based on our EVLA upper limit at $\Delta t \approx 1.2$ years.
We note that this model
accommodates the full transition from relativistic to non-relativistic
evolution. We built a collection of model light-curves spanning
parameter range, $A_*\approx [0.01-100]$ and $E\approx
[10^{49}-10^{52}]$~erg, maintaining the assumption of $p=3$ and
$\epsilon_e=\epsilon_B=0.1$. Here we adopted a jet opening angle of
$\theta_j=5^o$ and an off-axis viewing angle of $\theta_{\rm oa}=90^o$
(the most conservative scenario). We note that \cite{vanEerten11a} have developed off-axis GRB afterglow lightcurve models based on hydrodynamic simulations that reproduce the semi-analytic models presented here to within factors of a few \citep[see also][]{vanEerten10}.

\begin{figure}
\plotone{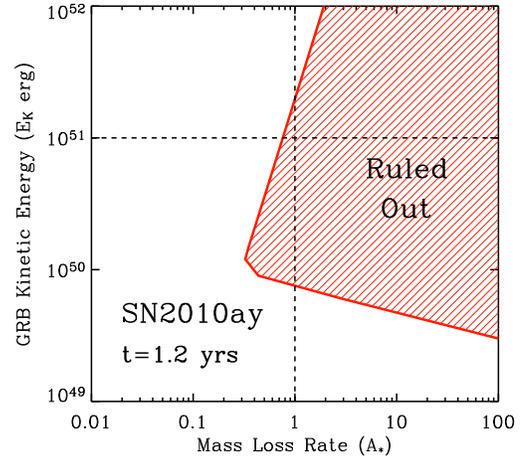}
\caption{Our EVLA observations at $\Delta t\approx 1.2$ years after explosion constrain the properties of a possibly associated off-axis GRB jet. Using our semi-analytic model as described in \S\ref{sec:offaxis}, we assume partition fractions of $\epsilon_e=\epsilon_B=0.1$, $\theta_j=5$ degrees, $p=2.5$, and an off-axis viewing angle of $\theta_{oa}=90$ degrees. We are able to exclude the region of $E_K-A_*$ parameter space (red) associated with typical GRBs, i.e. $E_K=10^{51}$~erg and $A_*=1$ (dashed black lines).\label{fig:oa}}
\end{figure}

As shown in
Figure~\ref{fig:oa}, we are able to exclude the parameter
space associated with typical GRBs, i.e. $E\approx 10^{51}$~erg
(beaming corrected) and $A_*\approx 1$. GRBs with lower energies and
densities are better constrained using the trans-relativistic formalism
above. In conclusion, our radio follow-up of SN\,2010ay reveals no
evidence for a relativistic outflow similar to those observed in
conjunction with the nearest GRB-SNe, however a weak afterglow like
that seen from XRF\,060218 cannot be excluded.

\section{Constraints on an associated GRB}\label{res:noGRB}

Given the estimate of the explosion date we have derived
(\S\ref{exp:lcmodel}), we have searched for gamma-ray emission that
may have been detected by satellites. No GRBs consistent with SN
2010ay were reported in the circulars of the Gamma-ray Coordinates
Network Circulars, but it is possible that bursts were detected below
the instrument triggering thresholds.

We next consulted the sub-threshold bursts from the Swift Burst Alert
Telescope (BAT; \citealt{Swift,SwiftBAT}) detected within the 6 days
surrounding the explosion date estimate.  We find that no gamma-ray
emission was detected within 0.5 degrees of the position of SN\,2010ay
by the BAT during this period. Given the sensitivity of the BAT, this
corresponds to an upper limit on the gamma-ray flux of
$\sim10^{-8}$~erg~s$^{-1}$~cm$^{-2}$ (15-150 keV). However, the field
of SN\,2010ay was in the field of view of the instrument for only 106
ksec during these 6 days, or $\sim20\%$ of the duration.

For complete temporal coverage, we have searched the records of the
interplanetary network (IPN), which is sensitive to bursts with
fluences down to $\sim6 \times 10^{-7}$~erg~cm$^{-2}$ (25-150 keV)
\citep[50\% efficiency limit,][]{IPN2}, and observes the entire sky

with a temporal duty cycle close to 100\%. An undetected,

sub-threshold burst should have a fluence below this limit. Between
2010 February 21 and 25, inclusive, a total of 12 bursts were detected
by the spacecraft of the IPN (Mars Odyssey, Konus-Wind, RHESSI,
INTEGRAL (SPI-ACS), Swift-BAT, Suzaku, AGILE, MESSENGER, and Fermi
(GBM)). Ten of them are confirmed bursts; they were observed by more
than one instrument on one or more spacecraft, and could be
localized. Two of them are unconfirmed bursts; they were observed by
one instrument on one spacecraft (Suzaku). The total area of the
localizations of the 10 confirmed bursts is $\sim0.58 \times 4
\pi$~sr. This implies that about 0.58 bursts can be expected to have
positions that are consistent with any given point on the sky simply
by chance (i.e. within the $3\sigma$ error region), and indeed {\it
none} of the bursts in this sample has a position consistent with the
SN position.

These non-detections imply upper-limits to the gamma-ray energy
($E_\gamma$) of a burst that may have been associated with SN
2010ay. The IPN non-detection indicates
$E_\gamma\lesssim6\times10^{48}$~erg (25-150 keV), while the BAT
non-detection indicates that the peak energy of the burst was
$\lesssim1\times10^{47}$~erg~s$^{-1}$ (15-150 keV) if the burst
occurred while in the field of view of the instrument.

We consider whether or not a hypothetical GRB associated with SN\,2010ay would have been detected by BAT or the IPN if it had characteristics similar to well-studied SN-associated GRBs.  The isotropic prompt energy release of long GRBs
is typically $E_{\gamma,\rm iso}\sim10^{52}$~erg, however the prompt
emission of the sub-energetic class of GRB-SNe can be several orders
of magnitude fainter \citep{Soderberg06Nat}. GRB 980425/SN 1998bw had
a peak gamma-ray luminosity of $\sim5\times10^{46}$~erg~s$^{-1}$
\citep[24-1820 keV][]{Galama98}, which is a factor of 2 below our BAT
limit, and $E_{\gamma,\rm iso}\sim5\times10^{47}$~erg \citep{Pian00},
more than a factor of five below our limit. Neither satellite should
have detected such a burst. In contrast, GRB 031203/SN\,2003lw had a
peak gamma-ray luminosity of $\sim1\times10^{49}$~erg~s$^{-1}$ (20-200
keV) and a total isotropic equivalent energy of $E_{\gamma,\rm
iso}=(4\pm1)\times10^{49}$~erg \citep{Sazonov04}, about two orders of
magnitude above the sensitivity of the BAT and twice the threshold of
the IPN, respectively. GRB 030329/2003dh was even more luminous, with
$E_{\gamma,\rm iso}\sim7\times10^{49}$~erg \citep{Hjorth03}. A
burst like GRB 031203 or 030329 should certainly have been detected by
IPN, or the BAT if it occurred while the field of SN\,2010ay was in the
field of view of the instrument. XRF\,060218/SN\,2006aj, an extremely
long-duration ($\Delta t\approx2000$ s) event, had a peak luminosity
observed by BAT of $\sim 10^{-8}$~erg cm$^{-2}$ ~s$^{-1}$ (15-150
keV), corresponding to a peak emission of
$\approx2.4\times10^{46}$~erg~s$^{-1}$ given the redshift of the burst
($z=0.033$), and a total isotropic equivalent energy of $E_{\gamma,\rm
iso}=(6.2\pm0.3)\times10^{49}$~erg \citep{Campana06}. If such a burst
was associated with SN\,2010ay, its peak emission may have been a
factor of four below the BAT sensitivity limit, although its total
isotropic energy emission is an order of magnitude larger than our IPN
limit for SN\,2010ay. Finally, the event whose host galaxy and
supernova properties seem most similar to SN\,2010ay, GRB 100316D/SN
2010bh had $E_{\gamma,\rm iso}\geq(5.9\pm0.5)\times10^{49}$~erg
\cite{Starling11} --- a full order of magnitude above our IPN limit.

Another possibility is that prompt emission associated with SN\,2010ay
may have been too soft to be detected by the BAT or IPN. For example,
the spectrum of XRF\,060218 rose to a peak at $0.3-10$~keV at $\sim985$ s
after triggering, then softened significantly thereafter. Even though
the the total emission of this burst is well above our IPN limit, it
may have escaped detection if it was similarly soft.

\section{Sub-solar Host Environment Metallicity} \label{res:metallicity}

We estimate the oxygen abundance of the host environment of SN\,2010ay
from the strong nebular emission line fluxes measured from our Gemini
spectrum (Table~\ref{tab:lines}).  At the distance of the host galaxy, the 1\asec\ Gemini slit width corresponds to a physical size of $1.3$~kpc.  The properties we infer for the explosion site of SN\,2010ay represent a luminosity-weighted average over this radius.

We employ several different oxygen abundance diagnostics in order to determine the metallicity of the host galaxy from its optical spectrum \citep[e.g.][]{Modjaz10}. From the O3N2 diagnostic of
\cite{PP04} (PP04), we derive a metallicity of $\log({\rm
O/H})=\GemPPohfourOthree{}$, or $Z\sim\GemPPohfoursol{}~Z_\odot$,
adopting the solar metallicity $\log({\rm O/H})_\odot+12=8.69$ from
\cite{asplund}. Using the N2 diagnostic of PP04, we find $\log({\rm
O/H})+12=\GemPPohfourNtwo{}$. Using the abundance diagnostic,
$R_{23}=\log([$\ion{O}{2}$] \lambda3727+[$\ion{O}{3}$]
\lambda\lambda4959,5007)/{{\rm H}\beta}$, we find
log(O/H)$+12_{\rm Z94}=\GemZninefour{}$ \citep{Zaritsky94} and
log(O/H)$+12_{\rm KD02}=\GemKDohtwo{}$ \citep{kewley02}. However, these
$R_{23}$-based estimates are more sensitive to flux-calibration and
reddening-correction. Moreover, there is a well-known bi-valued
relationship between $R_{23}$ and oxygen abundance. The value
$R_{23}=\GemRtwothree{}$ measured at the explosion site places it near the turnover
point, but we assume that it lies on the upper branch based on its [N
II]/[\ion{O}{2}] ratio, following \cite{KE08}. The metallicity values
we derive using the PP04 and KD02 diagnostics are approximately
equivalent given the offset that exists between these two diagnostics
\cite{KE08}. These measurements are similar to the values
reported by \cite{ModjazATel} (log(O/H)+12 [PP04,KD02]$=[8.2,8.4]$) for
the SN\,2010ay host galaxy. The statistical errors in our strong line
metallicity estimates are small ($< 0.01$ dex), as determined by
propagating the errors in the line flux measurement through the
abundance calculation. However, for example the representative
systematic error for the PP04 O3N2 abundance diagnostic is $\sim0.07$
dex, as determined by \cite{KE08} via comparison to other strong line
abundance indicators.

Fortunately, our detection of the weak [\ion{O}{3}] $\lambda 4363$
auroral line (S/N$\sim6$; Figure~3) allows us to derive an oxygen abundance via
the ``direct,'' $T_e$ method. We employ a methodology similar to that
used by, for example, \cite{Levesque10}. We first derive the electron
temperature ($T_e=\GemTe$~K) and density ($n_e=\GemNe$~cm$^{-3}$) from
the [\ion{O}{3}] and [S II] line ratios using the \texttt{temden} task
of the IRAF package \texttt{nebular} \citep{shaw94}, derive the $O^+$
temperature using the calibration of \cite{Garnett92}, and finally
estimate the O$^+$ and O$^{++}$ abundances following \cite{Shi06}. The
direct abundance, log(O/H)$+12=\GemDirect{}$, is in good agreement with the PP04 O3N2 value. The stated uncertainty reflects the
propagation of the uncertainties for the line flux
measurements. Indeed, the offset between these two diagnostics should
be very small at this metallicity \cite{KE08}.

We estimate the star formation rate (SFR) of the host galaxy using the
H$\alpha$ relation of \cite{Kennicutt98}.  After correcting for host
galaxy extinction, we measure the H$\alpha$ luminosity from our Gemini
spectrum (Table \ref{tab:lines}) and estimate
SFR$=\KSEsfr{}~M_{\odot}~{\rm yr}^{-1}$.

\subsection{Blue Compact Galaxy Host}
\label{host:jhu}

We compare the host galaxy of SN\,2010ay to the nearby galaxy
population of the SDSS spectroscopic survey. The physical properties
of the host galaxy, SDSS J123527.19+270402.7, are estimated in the
MPA/JHU catalog\footnote{http://www.mpa-garching.mpg.de/SDSS
\citep[described in ][and updated for SDSS
DR7]{Kauffmann03,Tremonti04,Brinchmann04,Salim07}}. The total
(photometric) galaxy stellar mass ($M_*$) is given as \SDSSMstar{},
the aperture-corrected SFR is \SDSSSFR{}, and the nuclear (fiber)
oxygen abundance (O/H$_o$) is log(O/H)$+12=$\SDSSOH{} on the scale of
\cite{Tremonti04} (T04). The specific star formation rate (SSFR) of
the host galaxy is then $\approx2.8^{+0.9}_{-0.4}$ Gyr$^{-1}$.  For
consistency, we consider these values of $M_*$, the oxygen abundance,
and the SFR for the host galaxy of SN\,2010ay when comparing to other
galaxies in the MPA/JHU catalog.

The oxygen abundance and SFR of the host galaxy of SN\,2010ay listed in
the MPA/JHU catalog are consistent with the values we derive in this
paper (see also \citealt{Kelly11}). The MPA/JHU catalog lists
metallicities on the T04 scale. Using the \cite{KE08} conversion to the
PP04 scale, the T04 metallicity estimate corresponds to a metallicity
of log(O/H)$+12=\SDSSOHppohfour{}$, which is $\sim 0.2$ dex higher
than the one we measure (log(O/H)$+12=\GemPPohfourOthree{}$). However,
there is a large ($\sim0.2$ dex) rms scatter between the PP04 O3N2 and
T04 diagnostics at the regime of log(O/H)$+12_{\rm PP04}\sim8.2$
\citep{KE08}.  The SFR in the MPA/JHU catalog is also in good
agreement with the value we estimate from the H$\alpha$ luminosity.
Although our estimate does not include an aperture correction, the
size of the Gemini slit ($1$\asec) should encompass
most of the star formation in the galaxy (Petrosian
$r=1.355$\asec; Table \ref{tab:SDSS}).

The mass to light ratio of the host galaxy of SN\,2010ay is low
compared to typical star-forming galaxies. To compare the host galaxy
to the general galaxy population, we select a subset of the MPA/JHU
catalog by requiring that estimates of $M_*$, SFR, and O/H$_o$ be
available and we remove AGN according to \cite{Kauffmann03}. We consider
\jhuAGNcutN{} starbursting galaxies following these constraints. The
host galaxy ranks in the
[\jhuAGNcutMp{},\jhuAGNcutSFRp{},\jhuAGNcutOHp{}] percentile in
[$M_*$,SFR,O/H$_o$] among these galaxies. Among the selected galaxies
with a stellar mass as low as the host galaxy\footnote{This subset is
selected such that the host galaxy of SN\,2010ay has the median mass:
$\jhuAGNMcutMdown{}<M_*<\jhuAGNMcutMup{}\times10^8$~M$_{\odot}$,
$N_{\rm sim}=\jhuAGNMcutN{}$.}, the median and standard deviation of
the B-band\footnote{We obtain B-band magnitudes by converting the
$k$-corrected $gri$ magnitudes given in the MPA/JHU catalog to BVR
magnitudes using the transformation of \cite{Blanton07}.} absolute
magnitude is \hostMBexpect{}~mag. With $M_B=$\HostAbsB{}~mag, the host of SN
2010ay is more luminous than other galaxies with a similar mass at the
$\hostMBdiscrep{}\sigma$ level. The discrepancy is due to the blue
color of the SN\,2010ay host galaxy, which indicates a stellar
population that is very young and therefore has a low stellar mass to
light ratio. Among the \jhuAGNCcutN{} galaxies in the MPA/JHU catalog
that meet the constraints above and have a color similar to the host
of SN\,2010ay (\jhuAGNCcutURcol{}, from SDSS fiber magnitudes), the
host galaxy has typical properties, with [$M_*$,SFR,O/H$_o$] in the
[\jhuAGNCcutMp,\jhuAGNCcutSFRp\jhuAGNCcutOHp] percentile.

Based on these properties, we classify the host galaxy of SN\,2010ay as a luminous Blue Compact Galaxy (BCG). BCGs span a large range in luminosity ($-21<M_B<-12$, where luminous BCGs have $M_B<-17$), but are distinguished by their blue colors ($B-V<0.45$), high SFR ($1<{\rm SFR}<20~M_\odot~{\rm yr}^{-1}$), and low metallicity ($Z_\odot/50<Z<Z_\odot/2$; \citealt{Kunth00,Kong02}). The host galaxy of SN\,2010ay has a luminosity ($M_B=$\HostAbsB{}), color ($B-V=0.11\pm0.07$), SFR (\SDSSSFR{}), and metallicity ($Z\sim\GemPPohfoursol{}~Z_\odot$) consistent with all these ranges.\footnote{We note that a large fraction of luminous BCGs show evidence for disturbed morphologies or interaction with close companions \citep{Garland04,LopezSanchez06}, but we do not see evidence for a companion at the depth of SDSS images of the host galaxy of SN\,2010ay.}

\subsection{Comparison to SNe~Ic-BL and GRB-SNe Host Galaxies}
\label{host:compare}

Our measurement of the metallicity from the Gemini spectrum indicates that the explosion site of SN\,2010ay is $\sim0.5~(0.2)$ dex lower in metallicity than the median SNe~Ic (Ic-BL) in the sample of \cite{Modjaz10}. In that sample, the median PP04 O3N2 metallicity measured at the explosion site of SNe~Ic is log(O/H)$+12\approx8.7$ and for Ic-BL is $\approx8.4$ dex, for 12 and 13 objects, respectively. If instead the KD02 metallicity is used, the median of the sample is $\approx8.9$ dex for SNe~Ic (13 objects) and $\approx8.7$ dex for Ic-BL (15 objects), so the abundance of the SN\,2010ay host galaxy is similarly low compared to the median.

The metallicity of the environment of SN\,2010ay is more similar to previously-studied nearby GRB-SN progenitors. A metallicity identical to our measurement was measured at the explosion site of SN\,2010bh \citep{Levesque11}: log(O/H)$+12=8.2$. In the survey of \cite{Levesque10}, and adding the measurement for SN\,2010bh, the GRB-SNe host galaxies have an average and standard deviation PP04 O3N2 metallicity of log(O/H)$+12=8.1\pm0.1$ on the PP04 scale, which is consistent with the SN\,2010ay environment. Among the 17 LGRB host galaxies surveyed in \cite{Savaglio09} the average metallicity is somewhat lower, $1/6~Z_\odot$ or log(O/H)$+12\sim7.9$, but these are at an average redshift of $z\sim0.5$ that is much higher than SN\,2010ay.

This evidence suggests that the host galaxy of SN\,2010ay has chemical properties more consistent with LGRBs/GRB-SNe than SNe~Ic-BL without associated GRBs; however, selection effects may mitigate this discrepancy.  SNe found in targeted surveys of luminous galaxies have host galaxy properties biased towards higher metallicities, due to the luminosity-metallicity ($L-Z$) relation \citep{Tremonti04}.  LGRBs are found in an untargeted manner through their gamma-ray emission and therefore are not biased by this relation.

SN\,2010ay joins a growing list of SNe~Ic-BL that have been discovered in low metallicity host galaxies.  Given the systematic uncertainty in strong line oxygen abundance diagnostics ($\sim0.07$ dex), we will consider host galaxies with metallicity log(O/H)$_{\rm PP04}+12<8.3$ ($Z\lesssim0.4Z_\odot$) to be in the low-metallicity regime of SN\,2010ay.  Among the 15 SNe~Ic-BL (9 discovered by untargeted searches) in the surveys of \cite{Modjaz08} and \cite{Modjaz10}, 4 were found in low metallicity environments: SN [2007eb,2007qw,2005kr,2006nx] at log(O/H)$_{\rm PP04}+12=$[8.26,8.19,8.24,8.24].  All of these SNe were discovered by untargeted searches.  \cite{Young10} measure the metallicity of the host galaxy of the broad-lined Ic SN\,2007bg to be log(O/H)$_{\rm PP04}+12=8.18$, although this SN has lightcurve and spectral properties that distinguish it from normal SNe~Ic-BL (\S\ref{obs:vel}).  Furthermore, \cite{Arcavi10} find that SNe~Ic-BL are more common in dwarf ($M_r\geq-18$) host galaxies, which the authors attribute to a preference for lower metallicities.  

The star formation properties of the host galaxy of SN\,2010ay also
resemble the host galaxies of LGRBs. If we consider those galaxies in
the MPA/JHU catalog with masses similar to the host galaxy of SN
2010ay (as defined above), then the median SFR and O/H$_o$ of these
galaxies is $\jhuAGNMcutSFRmed~M_\odot$~yr$^{-1}$ and
log(O/H$_o$)$+12=\jhuAGNMcutOHmed{}$, respectively. The host galaxy of
SN\,2010ay is in the [\jhuAGNMcutSFRp{},\jhuAGNMcutOHp{}] percentile
for [SFR,O/H$_o$] among these galaxies. This indicates that, while the
host galaxy of SN\,2010ay falls within $1\sigma$ of the
mass-metallicity ($M-Z$) relation for star-forming galaxies, its SFR is
extreme for its mass. The 39 LGRB host galaxies in the survey of
\cite{Savaglio09} are similarly low in mass and have high star
formation rates, with an average stellar mass of
$M_*\sim10^{9}~M_\odot$ and SSFR$\sim3.5$~Gyr$^{-1}$.

\begin{figure*}
\epsscale{.75}
\plotone{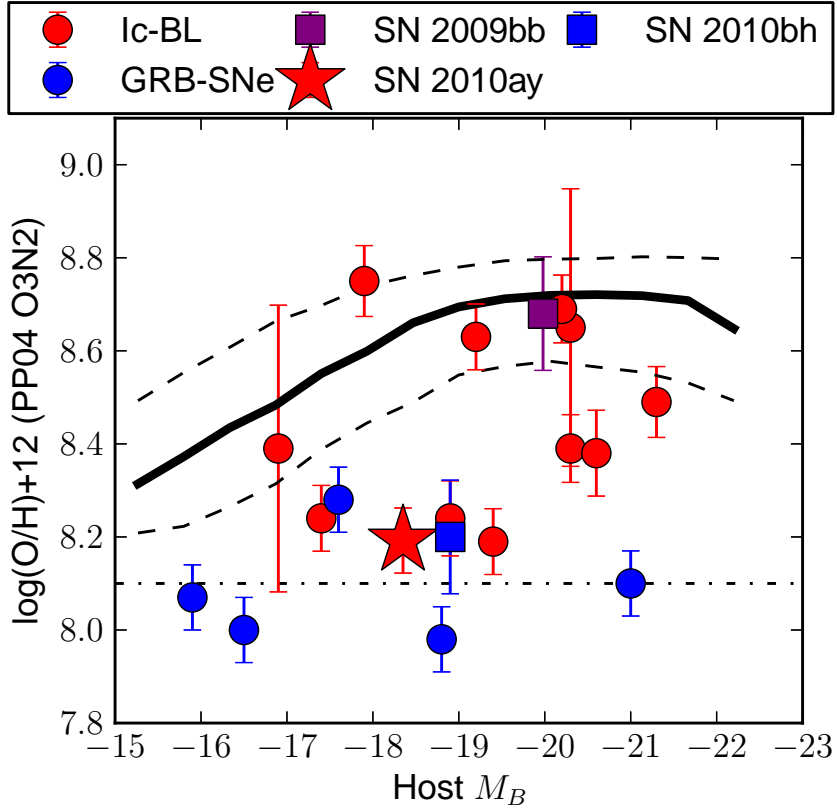}
\caption{A plot of host galaxy metallicity versus absolute B magnitude for SNe~Ic-BL (red) and engine-driven explosions (blue). The L-Z relation of nearby starforming galaxies is plotted as a solid line, with the 15th and 85th percentile boundaries of the galaxy distribution (dashed lines). Here we have transformed the $k$-corrected $gri$ magnitudes from the MPA/JHU catalog to B-band \citep{Blanton07} and converted the T04 metallicity values to the PP04 scale \citep{KE08}, for the purpose of comparing it to metallicity measurements for SN host galaxies in the literature. The dot-dashed horizontal line is the divider between GRB-SNe and SNe~Ic-BL host environments suggested by \cite{Modjaz08}. The host galaxy properties of GRB/SNe other than 2010ay are from the following references: \citealt{Starling11,Cano11} (2010bh), \citealt{Modjaz10} (Ic-BL), \citealt{Levesque102} (2009bb), and \citealt{Levesque10} (other GRB-SNe).  Errorbars illustrate measurement uncertainty, when published, plus a $0.07$~dex systematic uncertainty.\label{fig:tremonti}}
\end{figure*}

The host galaxy of SN\,2010ay falls below the $L-Z$ relation for nearby star-forming galaxies, as illustrated by Figure \ref{fig:tremonti}.  We have transformed the $k$-corrected $gri$ magnitudes from the MPA/JHU catalog to B-band \citep{Blanton07}. At the luminosity of the host galaxy of SN\,2010ay, the median metallicity and standard deviation of the SDSS galaxies on the T04 scale is log(O/H)$+12=$\jhuAGNMBcutOHmed{}; the host galaxy of SN\,2010ay falls in the \jhuAGNMBcutOHp{} percentile.  In other words, the host galaxy of SN\,2010ay is a $2\sigma$ outlier from the $L-Z$ relation. Similarly, \cite{Levesque10} and \cite{Han10} suggest that the host galaxies of LGRBs fall below the $L-Z$ relation as defined by normal star-forming galaxies, BCGs, and the host galaxies of Type Ic SNe.

\cite{Mannucci11} have explained the offset of LGRB host galaxies from the $M-Z$ relation as a preference for LGRBs to occur in host galaxies with high SFR, as characterized by the Fundamental Metallicity Relation (FMR) of \cite{Mannucci10} (see also \citealt{LaraLopez10}). Using the extended FMR for low mass galaxies from \citep{Mannucci11}, the host galaxy of SN\,2010ay should have a metallicity of log(O/H)$+12=\fmrOH{}$ given its stellar mass and SFR. The FMR is calibrated to the \cite{Nagao06} metallicity scale, which is similar to that of PP04 at this metallicity. Given the intrinsic scatter in the extended FMR on the order of $\sim0.05$~dex, this value is consistent with the PP04 value we measure from our Gemini spectrum: log(O/H)$+12=\GemPPohfourOthree$.  \cite{Kocevski11} similarly explain the offset of LGRB host-galaxies from the $M-Z$ relation as a SFR effect, but suggest that the long GRB host galaxies have even higher SFR than would be implied by the FMR. 

SN\,2010ay is an example of a SN~Ic-BL where the host galaxy is
consistent with the $M-Z$ relation for star-forming galaxies, but
deviates from the $L-Z$ relation due to its low stellar mass to light
ratio (\S\ref{host:jhu}).  Its $2\sigma$ discrepancy from the $L-Z$
relation would be hard to explain as a SFR rate effect alone because among
galaxies in the MPA/JHU catalog without AGN (as defined above) and
with $M_B$ within $0.1$~mag of the host galaxy of SN\,2010ay, the host
galaxy has a SFR in the \jhuAGNMBcutSFRp{} percentile ($<1\sigma$
discrepancy).

\section{Discussion} \label{res:phys}

\begin{deluxetable}{lcc}
\tablecaption{Comparison between SN\,2010ay and GRB 100316D/SN\,2010bh\label{tab:aybh}}
\tablewidth{0pt}
\tabletypesize{\scriptsize}
\tablehead{\colhead{Property} & \colhead{SN\,2010bh} & \colhead{SN\,2010ay}}
\startdata
\cutinhead{Host galaxy properties}
log(O/H)+12 \tablenotemark{a} & 8.2 & \GemPPohfourOthree \\ Redshift
(z) & 0.059 & \Gemz \\ $M_R$ & -18.5 & -18.94 \\
\cutinhead{Explosion properties}
SN type             & Ic-BL   & Ic-BL \\
$\vSi^{30}$ ($10^3 $~km~s$^{-1}$)\tablenotemark{b} & $\vthirtybh$     & $\vthirtyay$ \\
$M_R$        & $-18.60\pm0.08$     & $\Rpeak$ \\
$M_{\rm Ni}$ $(M_\odot)$            & $0.10\pm0.01$   & $\Mni{}$ \\
$M_{\rm ej}$ $(M_\odot)$            & 1.93-2.24   & $\gtrsim\Mej${} \\
$E_{K,51}$            & 12.0-13.9   & $\gtrsim\Ek{}$ \\
GRB energy ($E_{\rm iso}$,~erg) & $\gtrsim5.9\times10^{49}$ \tablenotemark{c} & $\lesssim6\times10^{48}$ \\
\enddata
\tablecomments{The observed properties of SN\,2010bh and its host galaxy are given by \cite{Chornock2010bh} and light curve modelling was performed by \cite{Cano11}. The properties of SN\,2010ay are derived in this paper.}
\tablenotetext{a}{The oxygen abundance derived from the PP04 O3N2 metallicity diagnostic, as discussed in \S\ref{res:metallicity}.}
\tablenotetext{b}{The absorption velocity at 30 days after explosion, as measured from the \ion{Si}{2}~$\lambda6355\rm \AA$ feature in \S\ref{res:snprop}.}
\tablenotetext{c}{The lower limit of the total isotropic energy release estimated by \cite{Starling11}.}

\end{deluxetable}

SN\,2010ay has all the hallmark features associated with GRB-SNe, and
yet we find no evidence of a relativistic explosion to sensitive limits. We are able to
place constraints on the energy, density, velocity, progenitor mass-loss rate,
and gamma-ray flux of any GRB that may have been associated with
it. In particular, we may rule out the association of a
GRB that looks similar to any spectroscopically confirmed GRB-SN to
date, except for XRF\,060218.

The low metallicity of the host environment of SN\,2010ay may be suitable for GRB jet formation in the ``collapsar'' model, but our observations strongly constrain any relativistic outflow (\S\ref{res:evla} \& \ref{res:noGRB}).  In \cite{MacFadyen99}, a high rate of rotation in the core of the progenitor is required to power a relativistic jet. A low metallicity is prescribed to suppress the line driven winds that would deprive the core of angular momentum. Apparently supporting this model, \cite{Stanek06} found that the isotropic prompt energy release of the GRB-SNe decreases steeply with metallicity, and other surveys have found observational evidence for the preferential occurrence of GRB-SNe in low-metallicity host galaxies \citep{Fynbo03,Prochaska04,Sollerman05,Modjaz06,Wiersema07,Christensen08,Modjaz08,Levesque10,Chornock2010bh,Starling11}. Challenging this view is the recent discovery of SN\,2009bb, a broad-lined, engine-driven Type-Ic supernova found in a high-metallicity host environment \citep{scp+10,Levesque102,Pignata11}. In SN\,2010ay, we have found the opposite case -- a broad-lined Type Ic supernova found in a low-metallicity host environment, but without any indication (via either radio or gamma-ray emission) of a central engine. The existence of SNe 2009bb and 2010ay emphasizes that progenitor metallicity is not the key factor that distinguishes GRB-SNe from broad-lined SNe~Ic without associated relativistic ejecta.

\begin{figure*}
\plotone{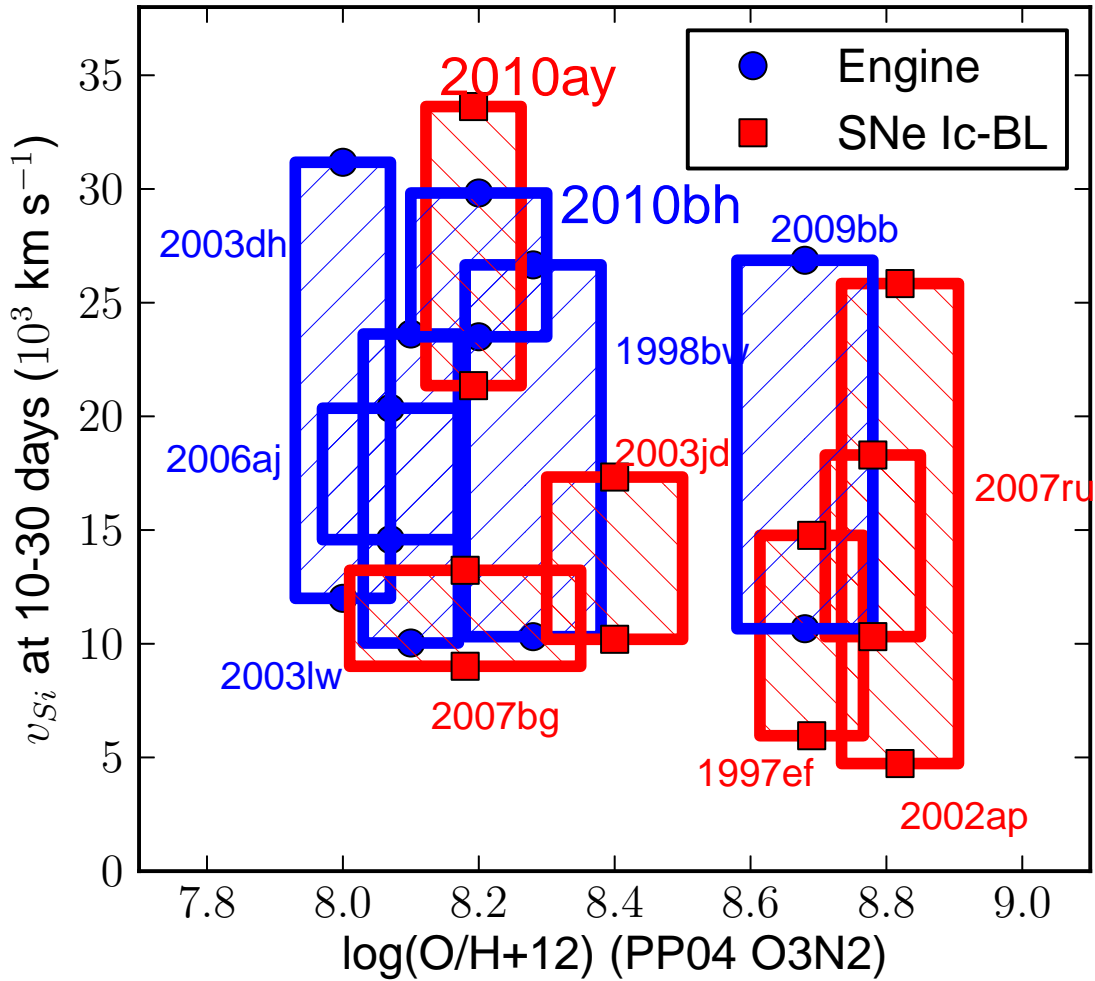}
\caption{SN absorption velocity, as traced by the \ion{Si}{2}~$\lambda6355\rm \AA$ feature, versus host galaxy oxygen abundance for SN\,2010ay and other Ic-BL (red) and engine-driven explosions (blue) from the literature (as in Figure \ref{fig:vc}). The range of velocities hatched for each object comes from the velocity at 10 days and at 30 days after explosion, according to the power law fits presented in Figure \ref{fig:vc}. The oxygen abundance measurements using the PP04 O3N2 diagnostic are from \cite{Levesque10} (GRB-SNe), \cite{Sahu09} (SN2007ru), \cite{Young10} (SN\,2007bg), and \cite{Modjaz10} (other SNe~Ic-BL). The range of oxygen abundance hatched reflects the error bars quoted in the literature (when stated) plus the $\sim0.07$ dex systematic error of the PP04 O3N2 diagnostic \citep{KE08}.\label{fig:velZ}}
\end{figure*}

We compare the absorption velocity of SNe~Ic-BL and engine-driven SNe (GRB-SNe and SN\,2009bb) to the metallicity of their host environments in Figure \ref{fig:velZ}. This comparison emphasizes the diversity of explosion and host galaxy properties observed in both engine-driven SNe and SNe~Ic-BL. The engine-driven SNe with the largest velocity gradients (2003dh, $\alpha=\vslopedh$; 2009bb, $\alpha=\vslopebw$) occur at metallicities different by a factor of 5 (2003dh, log(O/H)$+12=8.0$; 2009bb, log(O/H)$+12=8.7$ on the PP04 scale). Furthermore, the velocity gradient of engine-driven SNe seems to be uncorrelated with the velocity at late times; 2010bh ($\alpha=\vslopebh$) and 2006aj ($\alpha=\vslopeaj$) have similar velocity gradients, but 2010bh had velocities $\sim10,000$~km~s$^{-1}$ larger at 30 days after explosion. Among SNe~Ic-BL from low metallicity environments (log(O/H)$+12<8.5$), there is a large range in both the characteristic velocity and velocity gradient ($\vSi^{30}=[\vthirtyjd,\vthirtybg,\vthirtyay]$ and $\alpha=[\vslopejd,\vslopebg,\vslopeay]$ for SNe [2003jd,2007bg,2010ay]). For the three SNe~Ic-BL at higher metallicities, the characteristic velocities tend to be lower and the velocity gradients tend to be stronger ($\vSi^{30}=[\vthirtyef,\vthirtyap,\vthirtyru]$ and $\alpha=[\vslopeef,\vslopeap,\vsloperu]$ for SNe [1997ef,2002ap,2007ru]). However, a larger sample is needed to exclude the possibility of SNe~Ic-BL from super-solar metallicity environments that have high characteristic velocities or shallow velocity gradients.

The fact that a GRB was not detected in association with SN\,2010ay,
despite its similarities to the GRB-SNe, could indicate that the
relativistic jet produced by this explosion was ``suffocated'' before
it emerged from the progenitor star \citep{MWH01}. In this scenario,
the duration of the accretion event onto the newly-formed central
engine is short-lived and the jet post-breakout outflow is not
ultra-relativistic. In the process of being suffocated, the jet
transfers momentum to the ejecta such that the spectrum is broad lined
and the ejecta velocity is very high, even at late times, as we
observe (\S\ref{res:snprop}). However, the low host environment
metallicity we measure for SN\,2010ay, which is similar to GRB-SNe,
does not suggest high angular momentum loss that would help to weaken
the jet. Another alternative is that binary interaction plays a key role
in the commonality of relativistic outflows among SNe~Ic-BL.

Looking forward, additional SNe~Ic-BL in sub-solar metallicity host
environments will be found through current and future generations of
untargeted transient searches. Untargeted searches are required to
find supernovae from low-metallicity host environments, because
targeted searches only probe the highest metallicity galaxies due to
the luminosity-metallicity relationship
\citep{Modjaz10,Leloudas112}.

\section{CONCLUSIONS}
\label{sec:conc}

The optical photometric and spectroscopic, radio, and gamma-ray observations of SN\,2010ay presented here provide an example of a Type Ic-BL SN with explosion and host properties similar to the known GRB-SN SN\,2010bh. This object demonstrates that SNe in low-metallicity environments with high-velocity ejecta are not necessarily accompanied by the traditional signature of radio emission associated with long-lived relativistic jets. The existence of SN\,2010ay and SN\,2009bb (a central-engine driven event from a high-metallicity host environment) indicates that progenitor metallicity may not be the key factor that distinguishes GRB-SNe from normal broad-lined Type Ic supernovae.

We conclude that:

\begin{enumerate}
\item Pre-discovery imaging of the SN\,2010ay from the \PS\ $3\pi$ survey allows us to tightly constrain the early-time light-curve of SN\,2010ay (see Figure \ref{fig:lc}) and explosion date (\texp), allowing us to search for gamma-ray emission that may have been associated with the explosion. By fitting the template SN Ib/c light-curve of \cite{Drout10}, we derive an R-band peak absolute magnitude of $\Rpeak{}$~mag -- making SN\,2010ay among the most luminous SNe Ib/c ever observed. This peak magnitude suggests that a large mass of nickel, $M_{\rm Ni}\sim\Mni{}~M_\odot$, has been synthesized. We estimate a ratio of $M_{\rm Ni}$ to $M_{\rm ej}$ that is $\sim2\times$ larger than in known GRB-SNe.
\item Spectroscopy (see Figure \ref{fig:spectra}) at the explosion site in the host galaxy of SN\,2010ay indicates that the host environment of the progenitor star had a significantly sub-solar metallicity ($Z\sim \GemPPohfoursol~Z_\odot$), similar to the host environments of known GRB-SNe progenitors.
\item The Type Ic-BL SN\,2010ay strongly resembles the GRB-SN\,2010bh, particularly in light of its unusually high absorption velocities at late times ($\vSi\approx$\VelWhtTonrySiA{}$\times10^3$~km~s$^{-1}$ at \tpeakWHT{} days after peak) and low metallicity host environment. The comparison between these two SNe is summarized in Table \ref{tab:aybh}.
\item Non-detections in late-time EVLA radio observations of the SN rule out the association of a GRB of the nature of the spectroscopically-confirmed GRB-SNe, except for the radio afterglow associated with XRF\,060218. Our radio observations imply limits on the velocity, energy, and density of any associated relativistic jet and the mass loss rate of the progenitor (see Figures \ref{fig:radio1}, \ref{fig:ev_sn2010ay}, \ref{fig:ed_sn2010ay}, and \ref{fig:oa}). Additionally, no coincident gamma-ray emission was detected by satellites: the non-detection by the interplanetary network indicates $E_\gamma\lesssim6\times10^{48}$~erg, while the non-detection by the Swift BAT indicates that the peak energy of the burst was $\lesssim1\times10^{47}$~erg~s$^{-1}$ if the burst occurred during the $\sim20$\% of the explosion window when it was in the field of view of the instrument. This rules out associated prompt emission similar to that of GRBs 031203, 030329, or 100316D, but not GRBs 980425 or 060218.

\end{enumerate}

The pre-discovery imaging of SN\,2010ay demonstrates the capability of the untargeted PS1 survey for identifying and monitoring
exotic transients, not only in its high-cadence Medium-Deep Fields,
but also in the all-sky $3\pi$ survey.  Additional detections and
multi-wavelength follow-up observations of SNe~Ic-BL will help to
illuminate the role that ejecta velocity and progenitor metallicity
play in the GRB-SNe connection.

\acknowledgements
\label{sec:ackn}

The authors would like to thank the referee for useful comments and Andrew Drake, Angel L{\'o}pez-S{\'a}nchez, Andrew MacFadyen, Filippo Mannucci, and Maryam Modjaz for helpful discussions during the drafting of this paper. Additionally,
we thank the following members of the PS1 builders team: William
Burgett.  L.C. is a Jansky Fellow.  R.J.F. is a Clay Fellow.  E.M.L. is an Einstein fellow.

We are grateful to the following people for their assistance with the
IPN data: C. Meegan (Fermi GBM), K. Yamaoka, M. Ohno, Y. Hanabata,
Y. Fukazawa, T. Takahashi, M. Tashiro, T. Murakami, and K. Makishima
(Suzaku WAM), J. Goldsten (MESSENGER), S. Barthelmy, J. Cummings,
H. Krimm, and D. Palmer (Swift BAT), R. Aptekar, V. Pal'shin,
D. Frederiks, and D. Svinkin (Konus Wind), X. Zhang, and A. Rau
(INTEGRAL SPI-ACS), and I. G. Mitrofanov, D. Golovin, M. L. Litvak,
A. B. Sanin, C. Fellows, K. Harshman, H. Enos, and R. Starr
(Odyssey). The Konus-Wind experiment is supported in the Russian
Federation by RFBR grant 09-02-00166a. KH acknowledges NASA support
for the IPN under the following grants: NNX07AR71G (MESSENGER),
NNX08AN23G and NNX09AO97G (Swift), NNX08AX95G and NNX09AR28G
(INTEGRAL), NNX09AU03G (Fermi), and NNX09AV61G (Suzaku).

The PS1 Surveys have been made possible through contributions of the
Institute for Astronomy, the University of Hawaii, the Pan-STARRS
Project Office, the Max-Planck Society and its participating
institutes, the Max Planck Institute for Astronomy, Heidelberg and the
Max Planck Institute for Extraterrestrial Physics, Garching, The Johns
Hopkins University, Durham University, the University of Edinburgh,
Queen's University Belfast, the Harvard-Smithsonian Center for
Astrophysics, and the Las Cumbres Observatory Global Telescope
Network, Incorporated, the National Central University of Taiwan, and
the National Aeronautics and Space Administration under Grant
No. NNX08AR22G issued through the Planetary Science Division of the
NASA Science Mission Directorate.

The National Radio Astronomy Observatory is a facility of the National
Science Foundation operated under cooperative agreement by Associated
Universities, Inc.

This work was supported by the National Science Foundation through a
Graduate Research Fellowship provided to NES.

Based on observations obtained at the Gemini Observatory, which is
operated by the Association of Universities for Research in Astronomy
(AURA) under a cooperative agreement with the NSF on behalf of the
Gemini partnership: the National Science Foundation (United States),
the Science and Technology Facilities Council (United Kingdom), the
National Research Council (Canada), CONICYT (Chile), the Australian
Research Council (Australia), CNPq (Brazil) and CONICET (Argentina)

{\it Facilities:} \facility{PS1}, \facility{Gemini:North} (GMOS-N), \facility{EVLA}, \facility{ING:Herschel}, \facility{Swift}

\bibliographystyle{fapj.bst}

\end{document}